\documentclass[usenatbib, useAMS]{mn2e}

\usepackage{graphicx}
\usepackage{txfonts}

\newcommand{\pd}{\ensuremath{ P \left( D \right) }}
\newcommand{\mum}{\ensuremath{ \mu }m}

\begin{document}
\title[HerMES \pd\ Fluctuation Analysis]
 {HerMES: Deep Galaxy Number Counts from a \pd\ Fluctuation Analysis of SPIRE
   Science Demonstration Phase Observations}
\author[J.~Glenn et al.]
{\parbox{\textwidth}{J.~Glenn,$^{1}$\thanks{E-mail: 
  \texttt{jason.glenn@.colorado.edu}}
A.~Conley,$^{1}$
M.~B{\'e}thermin,$^{2}$
B.~Altieri,$^{3}$
A.~Amblard,$^{4}$
V.~Arumugam,$^{5}$
H.~Aussel,$^{6}$
T.~Babbedge,$^{7}$
A.~Blain,$^{8}$
J.~Bock,$^{8,9}$
A.~Boselli,$^{10}$
V.~Buat,$^{10}$
N.~Castro-Rodr{\'\i}guez,$^{11,12}$
A.~Cava,$^{11,12}$
P.~Chanial,$^{7}$
D.L.~Clements,$^{7}$
L.~Conversi,$^{3}$
A.~Cooray,$^{4,8}$
C.D.~Dowell,$^{8,9}$
E.~Dwek,$^{13}$
S.~Eales,$^{14}$
D.~Elbaz,$^{6}$
T.P.~Ellsworth-Bowers,$^{1}$
M.~Fox,$^{7}$
A.~Franceschini,$^{15}$
W.~Gear,$^{14}$
M.~Griffin,$^{14}$
M.~Halpern,$^{16}$
E.~Hatziminaoglou,$^{17}$
E.~Ibar,$^{18}$
K.~Isaak,$^{14}$
R.J.~Ivison,$^{18,5}$
G.~Lagache,$^{2}$
G.~Laurent,$^{19}$
L.~Levenson,$^{8,9}$
N.~Lu,$^{8,20}$
S.~Madden,$^{6}$
B.~Maffei,$^{21}$
G.~Mainetti,$^{15}$
L.~Marchetti,$^{15}$
G.~Marsden,$^{16}$
H.T.~Nguyen,$^{9,8}$
B.~O'Halloran,$^{7}$
S.J.~Oliver,$^{22}$
A.~Omont,$^{23}$
M.J.~Page,$^{24}$
P.~Panuzzo,$^{6}$
A.~Papageorgiou,$^{14}$
C.P.~Pearson,$^{25,26}$
I.~P{\'e}rez-Fournon,$^{11,12}$
M.~Pohlen,$^{14}$
D.~Rigopoulou,$^{25,27}$
D.~Rizzo,$^{7}$
I.G.~Roseboom,$^{22}$
M.~Rowan-Robinson,$^{7}$
M.~S\'anchez Portal,$^{3}$
B.~Schulz,$^{8,20}$
Douglas~Scott,$^{16}$
N.~Seymour,$^{24}$
D.L.~Shupe,$^{8,20}$
A.J.~Smith,$^{22}$
J.A.~Stevens,$^{28}$
M.~Symeonidis,$^{24}$
M.~Trichas,$^{7}$
K.E.~Tugwell,$^{24}$
M.~Vaccari,$^{15}$
I.~Valtchanov,$^{3}$
J.D.~Vieira,$^{8}$
L.~Vigroux,$^{23}$
L.~Wang,$^{22}$
R.~Ward,$^{22}$
G.~Wright,$^{18}$
C.K.~Xu$^{8,20}$ and
M.~Zemcov$^{8,9}$}\vspace{0.4cm}\\
\parbox{\textwidth}{$^{1}$Dept. of Astrophysical and Planetary Sciences, CASA 389-UCB, University of Colorado, Boulder, CO 80309, USA\\
$^{2}$Institut d'Astrophysique Spatiale (IAS), b\^atiment 121, Universit\'e Paris-Sud 11 and CNRS (UMR 8617), 91405 Orsay, France\\
$^{3}$Herschel Science Centre, European Space Astronomy Centre, Villanueva de la Ca\~nada, 28691 Madrid, Spain\\
$^{4}$Dept. of Physics \& Astronomy, University of California, Irvine, CA 92697, USA\\
$^{5}$Institute for Astronomy, University of Edinburgh, Royal Observatory, Blackford Hill, Edinburgh EH9 3HJ, UK\\
$^{6}$Laboratoire AIM-Paris-Saclay, CEA/DSM/Irfu - CNRS - Universit\'e Paris Diderot, CE-Saclay, pt courrier 131, F-91191 Gif-sur-Yvette, France\\
$^{7}$Astrophysics Group, Imperial College London, Blackett Laboratory, Prince Consort Road, London SW7 2AZ, UK\\
$^{8}$California Institute of Technology, 1200 E. California Blvd., Pasadena, CA 91125, USA\\
$^{9}$Jet Propulsion Laboratory, 4800 Oak Grove Drive, Pasadena, CA 91109, USA\\
$^{10}$Laboratoire d'Astrophysique de Marseille, OAMP, Universit\'e Aix-marseille, CNRS, 38 rue Fr\'ed\'eric Joliot-Curie, 13388 Marseille cedex 13, France\\
$^{11}$Instituto de Astrof{\'\i}sica de Canarias (IAC), E-38200 La Laguna, Tenerife, Spain\\
$^{12}$Departamento de Astrof{\'\i}sica, Universidad de La Laguna (ULL), E-38205 La Laguna, Tenerife, Spain\\
$^{13}$Observational  Cosmology Lab, Code 665, NASA Goddard Space Flight  Center, Greenbelt, MD 20771, USA\\
$^{14}$Cardiff School of Physics and Astronomy, Cardiff University, Queens Buildings, The Parade, Cardiff CF24 3AA, UK\\
$^{15}$Dipartimento di Astronomia, Universit\`{a} di Padova, vicolo Osservatorio, 3, 35122 Padova, Italy\\
$^{16}$Department of Physics \& Astronomy, University of British Columbia, 6224 Agricultural Road, Vancouver, BC V6T~1Z1, Canada\\
$^{17}$ESO, Karl-Schwarzschild-Str. 2, 85748 Garching bei M\"unchen, Germany\\
$^{18}$UK Astronomy Technology Centre, Royal Observatory, Blackford Hill, Edinburgh EH9 3HJ, UK\\
$^{19}$Southwest Research Institute, Boulder, Colorado 80302, USA\\
$^{20}$Infrared Processing and Analysis Center, MS 100-22, California Institute of Technology, JPL, Pasadena, CA 91125, USA\\
$^{21}$School of Physics and Astronomy, The University of Manchester, Alan Turing Building, Oxford Road, Manchester M13 9PL, UK\\
$^{22}$Astronomy Centre, Dept. of Physics \& Astronomy, University of Sussex, Brighton BN1 9QH, UK\\
$^{23}$Institut d'Astrophysique de Paris, UMR 7095, CNRS, UPMC Univ. Paris 06, 98bis boulevard Arago, F-75014 Paris, France\\
$^{24}$Mullard Space Science Laboratory, University College London, Holmbury St. Mary, Dorking, Surrey RH5 6NT, UK\\
$^{25}$Space Science \& Technology Department, Rutherford Appleton Laboratory, Chilton, Didcot, Oxfordshire OX11 0QX, UK\\
$^{26}$Institute for Space Imaging Science, University of Lethbridge, Lethbridge, Alberta, T1K 3M4, Canada\\
$^{27}$Astrophysics, Oxford University, Keble Road, Oxford OX1 3RH, UK\\
$^{28}$Centre for Astrophysics Research, University of Hertfordshire, College Lane, Hatfield, Hertfordshire AL10 9AB, UK}}

\maketitle
\clearpage

\begin{abstract}
Dusty, star forming galaxies contribute to a bright, currently
unresolved cosmic far-infrared background.  Deep {\it Herschel}-SPIRE
images designed to detect and characterize the galaxies that comprise
this background are highly confused, such that the bulk lies below the
classical confusion limit.  We analyze three fields from the HerMES
programme in all three SPIRE bands (250, 350, and 500 \mum );
parameterized galaxy number count models are derived to a depth of
$\sim 2$ mJy/beam, approximately 4 times the depth of previous
analyses at these wavelengths, using a \pd\ (probability of
deflection) approach for comparison to theoretical number count
models.  Our fits account for 64, 60, and 43 per cent of the
far-infrared background in the three bands.  The number counts are
consistent with those based on individually detected SPIRE sources,
but generally inconsistent with most galaxy number counts
models, which generically overpredict the number of bright galaxies
and are not as steep as the \pd-derived number counts.  Clear evidence
is found for a break in the slope of the differential number counts at
low flux densities.  Systematic effects in the \pd\ analysis are
explored.  We find that the effects of clustering have a small impact
on the data, and the largest identified systematic error arises from
uncertainties in the SPIRE beam.
\end{abstract}

\section{Introduction}
\nobreak 
The cosmic far-infrared background (hereafter CFIRB) provides
unique information on the history of energy injection in the Universe
by both star formation and active galactic nuclei. First detected by
the {\it COBE} satellite \citep{pug96,Fixsen:1998}, the CFIRB contains a large
amount of energy, indicating that the total luminosity from thermal
dust emission is comparable to the integrated UV/optical energy output
of galaxies \citep{guid97}.

Galaxy surveys, both from the ground (with SCUBA, LABOCA, Bolocam,
AzTEC, and MAMBO at 850 \mum , 870 \mum, 1.1 mm, 1.1 mm, and 1.3 mm,
respectively) and from space using {\it IRAS} (at 12, 25, 60 and 100
\mum), {\it ISO} (at 15, 90, and 170 \mum ), and {\it Spitzer} (at 3.6
to 160 \mum), found high number counts compared to non-evolving galaxy
number counts models.  This implied that strong evolution of the source
populations must have occurred, challenging contemporary galaxy
evolution models \citep{Saunders:90,scott02,lag03}.  Deeper
number counts test galaxy formation models more severley.  By
stacking {\it Spitzer} MIPS 24~\mum\ sources, at least 80\% of the
CFIRB was resolved at 70~\mum\ and 65\% at
160~\mum\ \citep{dol06,Bethermin:2010a}.  A small fraction (10-20\%)
has been resolved in the submillimeter in blind sky surveys from
ground-based observatories, but it is possible to go deeper by taking
advantage of gravitational lensing.  At 850~\mum\ this approach has
resolved 60\% or more of the background in small fields
\citep{smail,Zemcov:2010}.

A \pd\ -- probability of deflection -- analysis of Bolocam
observations of the Lockman Hole \citep{mal05} demonstrated that a
fluctuation analysis can provide more stringent constraints on source
number counts than those derived by extracting individual sources, for
which the threshold must be set high enough to ensure a minority of
false detections.  \pd\ techniques were first developed for
application to radio observations \citep{scheuer57}, but have since
been widely applied to other regimes.  \pd\ was used to account for
the majority of the X-ray background long prior to the availability of
sufficiently deep imaging to resolve individual sources
\citep{Barcons:1994}, to extend deep infrared counts
\citep{Oliver:1997}, and in the sub-mm to SCUBA \citep{Hughes:1998},
LABOCA \citep{Weiss:2009}, and AzTEC \citep{Scott:2010} data.  The
depth of a \pd\ analysis is set by the flux density at which the
number of sources per beam becomes large.  The resulting
contribution to the \pd\ becomes that of a Poissonian distribution
with a large mean, which becomes difficult to distinguish from the
nearly-Gaussian instrumental noise.  An often-used rule of thumb for
the maximum depth is one source per beam, but the precise limit
depends on the survey area, the shape of the underlying counts, and
how precisely the instrumental noise is known.  In practice, for
rapidly rising source counts at faint fluxes, this is considerably
deeper than the limits for a source-extraction approach.  Fluctuation
analyses are well-suited to determination of source number counts in
the case where the dynamic range of detected sources is not large
because of confusion.  Deep number counts are interesting because they
allow us to measure the sources responsible for the bulk of the CFIRB,
and because they probe intrinsically fainter galaxies which may have
better matching counterparts in the local Universe.

Recently, a \pd\ analysis was performed on 250, 350, and
500~\mum\ observations of a 10 $\mathrm{deg}^2$ field (GOODS-S) with a
0.8 $\mathrm{deg}^2$ deep inner region from the balloon-borne BLAST
telescope, using duplicate SPIRE detector technology \citep[][hereafter
  P09]{pat09}.  Differential number counts were estimated down to 20,
15, and 10 mJy in the three bands, respectively.  Below these
thresholds, upper limits were provided.  Combined with
24~\mum\ observations, \citet{Devlin:2009} concluded that a large
fraction ($> 1/2$) of the CFIRB comes from galaxies with $z>1.2$.
Also from BLAST observations, \citet{mar09} concluded that 24~\mum
-selected galaxies can account for the entire CFIRB based on a
stacking analysis.  These results confirm that fluctuation and
stacking analyses have substantial power in elucidating the sources of
the CFIRB.  Such techniques will also be necessary for SPIRE
observations because galaxy models predict that at the confusion limit
SPIRE is expected to resolve only a small fraction of the CFIRB
\citep{lag03, fer08}.  A recent source extraction-based analysis of
the SPIRE Science Demonstration Phase (SDP) data -- the same data used in
this paper -- directly resolved $15$, $10$, and $6$ per cent of the CFIRB
at 250, 350, and 500~\mum , respectively \citep{Oliver:2010b}.
At shorter wavelengths, \citet{Berta:2010} directly resolved
52 and 45 per cent of the CFIRB at 100 and 160 \mum\ using
{\it Herschel}-PACS SDP data.

\section{Data}
\label{sec:data}
\nobreak 
The observations used in this analysis were obtained with the SPIRE
instrument \citep{Griffin:2010} on the {\it Herschel} Space
Observatory \citep{Pilbratt:2010} as part of the HerMES
programme\footnote{http://www.hermes.sussex.ac.uk/} (Oliver et
al.\ 2010, in prep) during the SDP.  SPIRE observes simultaneously in
three passbands: 250, 350 and 500~\mum .  The on-orbit beam sizes,
including the effects of the scanning strategy, are 18.1, 25.2, and
36.6 arcseconds, respectively, with mean ellipticities of 7 - 12\%.
The calibration is based on observations of Neptune, and is described
in \citet{Swinyard:2010}.  Observations of five fields were obtained
during SDP, but only three are used in this analysis: GOODS-N,
Lockman-North, and Lockman-SWIRE.  Their properties are summarized in
table~\ref{tbl:data}.  The Lockman-North region are contained within
the shallower Lockman-SWIRE field.  The HerMES SDP fields omitted from
this analysis are: FLS, which was left out because it is the same
depth as the much-larger Lockman-SWIRE field and is significantly
contaminated by infrared cirrus, and Abell 2218, because the strong
lensing in this field complicates the interpretation of the background
number counts.

\begin{figure}
\centering
\includegraphics[width=9cm]{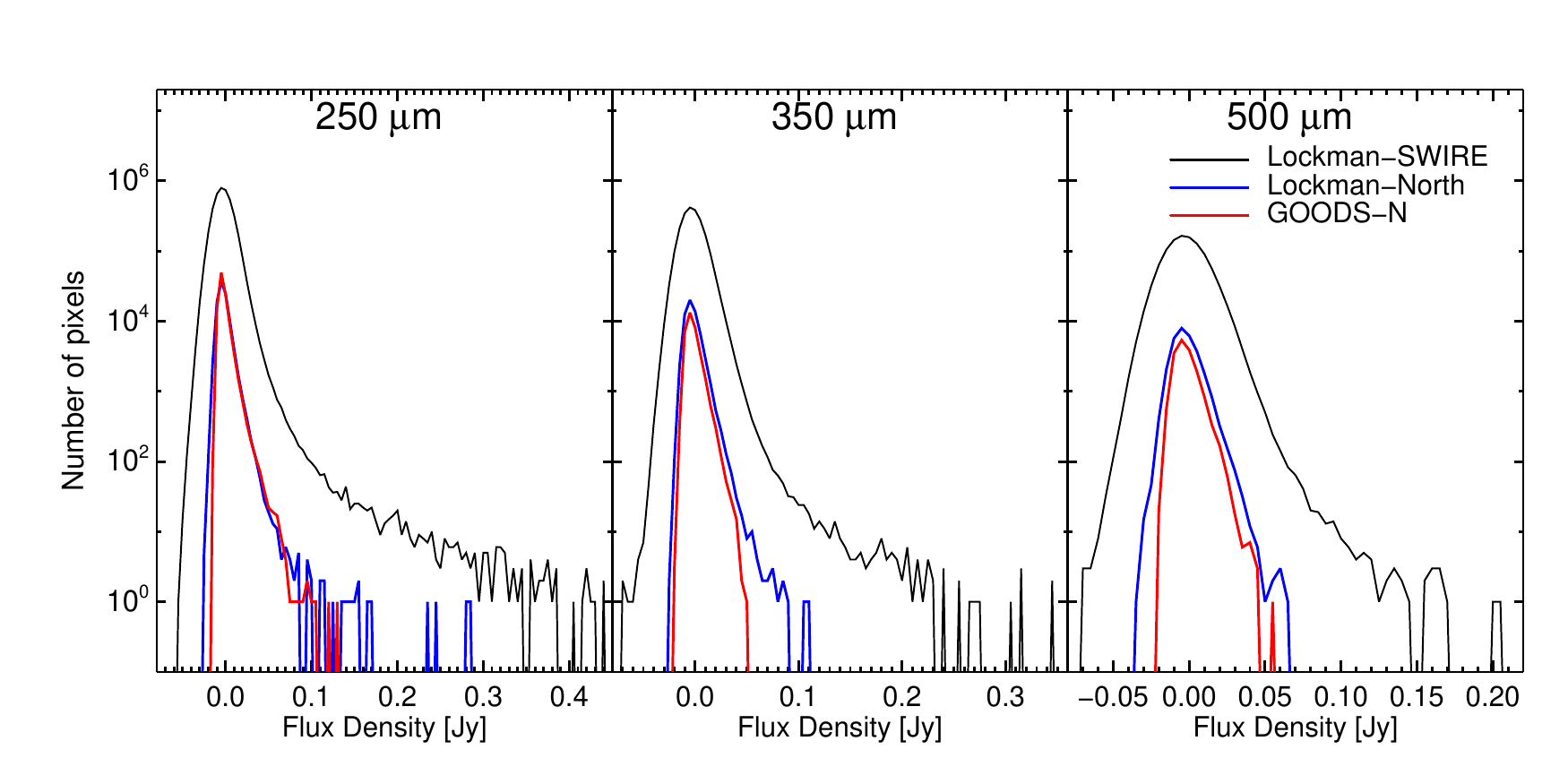}
\caption{Histogram of pixel flux densities for the three fields
  considered in this paper in 5 mJy bins.  The Lockman-SWIRE field is
  considerably shallower than the others, but it is the only field
  large enough to probe the bright end of the source distribution.
  The maps have been mean subtracted.  The binning here is for display
  purposes and does not correspond to the binning used in the actual
  analysis, which is much finer.  In all cases the pixel histograms
  show clear non-Gaussianity despite the Gaussian nature of the noise,
  indicating that a significant point-source contribution is
  present. \label{fig:histos} }
\end{figure}

The detector (bolometer) timelines were processed using the standard
SPIRE pipeline, which detects cosmic rays and removes instrumental
signatures and temperature drifts \citep{Dowell:2010}.  The maps were
produced using the SMAP package (Levenson et al.\ 2010, in prep) using
$1/3$ beam-FWHM pixels (6, 8 1/3, and 12 arcsec); this is a compromise
between adequately sampling the beam and maintaining even coverage
over the map.  Samples flagged as contaminated by cosmic-rays were
excluded.  Each map was masked to form an even-coverage region and was
mean subtracted.  In addition to the even-coverage mask, a small
amount of additional masking was required, as there are five resolved
sources in the Lockman-SWIRE field. These sources are relatively
bright, but are not the brightest in the field.  Since the
\pd\ formalism is based on unresolved point sources, we mask these
objects with a 2 arcmin circular mask, and then correct our final
number counts using the measured flux of each excluded source.  Our
instrumental noise estimates are based on the technique of
\citet{Nguyen:2010}, and are assumed to have 5\% uncertainty, which
represents only the uncertainty for a fixed calibration.  The overall
SPIRE calibration error is discussed in
\S~\ref{subsec:systematics}. The resulting pixel flux density
histograms are shown in figure~\ref{fig:histos}.

Smoothing the maps by the beam (via cross-correlation) is beneficial
for finding isolated point sources.  For BLAST observations,
\citet{Chapin:2010} show that the standard point-source-optimized
filter should be modified in the presence of confusion noise.
However, there is no guarantee this will benefit a
\pd\ analysis. Smoothing the map has the effect of broadening the
effective beam, which decreases the depth that the \pd\ can probe,
while also reducing (but correlating) the instrumental noise.  We
empirically determined if smoothing is beneficial for our analysis by
fitting simulated data with and without smoothing and comparing the
scatter in the recovered model parameters to the error estimates, and
found that it helps for all but our deepest map (GOODS-N); note that
our GOODS-N data are several times deeper (relative to the confusion
noise) than the deepest BLAST observations.  It is likely that some
amount of spatial de-convolution would be beneficial for the GOODS-N field,
but since this would significantly complicate the instrumental noise
properties of our maps, and hence require extensive testing, we do not
pursue de-convolution here, but do beam-smooth the two shallower
fields.  In addition, we have applied a 6 arcmin high-pass spatial
filter to our maps to reduce the effects of clustering on our results.
The motivation for this is discussed in \S~\ref{subsec:filtering}.

A \pd\ analysis is critically dependent on measurements of the
effective beam (which includes the effects of the map making and
reduction, as well as any smoothing applied).  Our beam map is based
on in-flight fine-scan (highly oversampled) observations of Neptune
with a large number of repeats and small offsets between each scan.
These observations allowed us to measure the beam with finer
resolution than our data maps to properly match the SPIRE calibration
(which is timestream rather than map based), and to build beam maps
for the individual bolometers.  We corrected these observations for
the relative motion of Neptune during the scans using the
HORIZONS\footnote{http://ssd.jpl.nasa.gov/?horizons} ephemeris
computation service at the orbit of {\it Herschel}.  The Neptune
observations are deep enough that the third Airy ring is clearly
detected for the array-averaged beams.  As discussed later
(\S~\ref{subsec:systematics}), beam uncertainties are our largest
identified systematic.

\begin{table*}
 \begin{minipage}{130mm}
 \centering
 \caption{HerMES SPIRE Science Demonstration Phase observations used
   in this paper}
 \begin{tabular}{@{}lllllllll@{}}
  \hline \hline
  Field & Size & RA & Dec & Scan Rate & 
    Repeats & $\sigma_{250}$ &
        $\sigma_{350}$ & $\sigma_{500}$ \\
        & $\mathrm{deg}^2$ & $\mathrm{deg}^2$ & $\mathrm{deg}^2$ & 
        arcsec/s & & mJy/beam &
        mJy/beam & mJy/beam \\
  \hline
  GOODS-N & 0.29 & 189.23 & 62.24 & 30 & 30 & 1.77 & 1.59 & 1.89 \\
  Lockman-North & 0.41 & 161.50 & 59.02 & 30 & 7 & 3.58 & 3.16 &
        4.41 \\
  Lockman-SWIRE & 13.6 & 162.0 & 58.11 & 60 & 2 & 9.47 & 8.47 &
        11.99 \\
  \hline
 \end{tabular}
 The $\sigma$ values for each band and field are the instrumental
 noise per pixel before any filtering or smoothing is applied.  The
 confusion noise (the signal in this analysis) is $\sim 6$ mJy/beam in
 all bands \citep{Nguyen:2010}.  \label{tbl:data}
\end{minipage}
\end{table*}

\section{Method}
\label{sec:method}
\nobreak 
We first describe the basic \pd\ framework, then discuss our
particular implementations and the filtering we have applied to limit
the effects of clustering.

\subsection{Description of \pd }
\label{subsec:pd}
\nobreak
If $dN/dS \! \left(S\right)$ is the differential number counts per solid
angle and $S$ the flux density, then the mean number density
of sources per unit solid angle with observed flux densities between 
$x$ and $x + dx$ is
\begin{equation}
 R\left( x \right) \, dx = 
 \int_{\Omega}\, \frac{dN}{dS} \! \left( \frac{x}{b} \right) 
       \, b^{-1} \, d\Omega\, dx
\end{equation}
where $b$ is the beam function (not necessarily peak normalized).
Ignoring clustering, the probability distribution of sources is
Poissonian.  The probability distribution function (pdf) for the
observed flux in each sky area unit (usually a map pixel) is the
convolution of the pdfs for each flux interval over all fluxes; this
quantity is called the \pd .  Rewriting the above in terms of
characteristic functions and denoting the inverse Fourier transform
by $F_{\omega}^{-1}$,
\begin{equation}
 \pd = F_{\omega}^{-1} \left[ \exp \left( \int_0^{\infty}\, R\left(x\right) 
    \exp \left(i \omega x\right) dx - \int_0^{\infty}\, R\left( x \right) dx 
     \right)
    \right] .
\end{equation}
The mean of the \pd\ is 
\begin{displaymath}
 \mu = \int x R \, dx = \int\, b\, d\Omega\, \int\, S\, dN/dS \! \left(S\right) 
   \, dS ,
\end{displaymath}
and the variance is
\begin{displaymath}
 \sigma^2_P = \int\, b^2 d\Omega\, \int\, S^2 dN/dS \! \left(S\right)\, dS.
\end{displaymath}
For real observations, the instrumental noise contribution must also
be included.  Our observations are not sensitive to the
mean flux in the maps.  Therefore, it is useful to subtract off the
mean of the \pd\ during construction.  Only in the case of very simple
models for $dN/dS$ combined with trivial beams is it possible to
compute \pd\ analytically -- an example is given in
\citet{scheuer57}, but even this is only valid for a restricted range
of parameters.  For an effective beam that is not strictly positive
(due to filtering, for example), the \pd\ is the convolution of the
individual \pd s for the positive beam and for the negative beam (P09).
Azimuthally averaging the beam does not preserve the \pd , so it
is necessary to use the full 2D beam map.

The log likelihood ($\log \mathcal{L}$) of a dataset relative to a
particular model is given (to within an irrelevant normalizing
constant) by
\begin{displaymath}
 \log \mathcal{L} = \sum_{i} \log P \left( D_i \right) ,
\end{displaymath}
where $D_i$ is the value of the $i^{\mathrm{th}}$ pixel.  Usually it is
more convenient to bin the data.  As long as the individual bins are
small compared to the width of the \pd , the two formulations are
practically equivalent.  Then
\begin{equation}
 \log \mathcal{L} = \sum_k h_k \log P \left( x_k \right),
\end{equation}
where $h_k$ is the number of pixels in the histogram bin centred at
flux density $x_k$. The alternative of using the $\chi^2$ as the fit
statistic under-weights bins with a small number of pixels in them
because the uncertainty in such a bin is not well modeled by
$\sqrt{h_k}$, and is not recommended.

This treatment assumes that different pixels are uncorrelated, which
is not true unless the beam is much smaller than a pixel.  A source at
one location will affect neighbouring values over an area about equal
to the area of the beam.  The result is that, if applied naively, fits
based on the above likelihood will underestimate the model errors.
Properly treating this effect requires developing the \pd\ formalism
in terms of multi-variate Poisson distributions, which is
computationally infeasible.  P09 recommend dividing the likelihood by
the beam area in pixels ($\mathcal{L} \mapsto \mathcal{L} / A_b$) in
order to correct for this effect, which amounts to approximately
correcting the likelihood for the number of independent samples in the
map. This is an ad hoc approach, but in the absence of a better
alternative, we have adopted a similar method.  However, based on
Monte-Carlo simulations of synthetic data sets, we find that this
correction factor is overly conservative, as discussed below.

Another approximation in the above treatment which is not valid for
real data is that the sources are not Poisson distributed due to
clustering.  Our approach to this effect is described in
\S~\ref{subsec:filtering}.

\subsection{Implementation}
\label{subsec:implementation}
\nobreak 
We have developed two independent \pd\ analysis packages and checked
them against each other.  Given the large number of parameters and the
non-linear nature of the problem, both make use of Markov Chain
Monte Carlo (MCMC) methods.  Overviews of MCMC methods can be found in
\citet{MacKay:02, Lewis:02, Dunkley:05}.  The important aspects of an
MCMC implementation are the burn-in criterion and the proposal
density.  The burn-in criterion is the rule used to determine
whether the fit has converged on the region of maximum likelihood.  Once
the fit has converged, subsequent steps are drawn from the posterior
probability of the model given the data, and only these steps are used
to measure the errors and values of the parameters.  The proposal
density is used to propose the next step in the Markov Chain from the
current step.  Any proposal density that can visit all valid
parameters is correct, but a well chosen density can dramatically
improve the efficiency of the fitting procedure.

The first code is written in IDL\footnote{Interactive Data Language:
  http://www.ittvis.com/ProductServices/IDL.aspx} and the burn-in
criterion is based on the power spectrum of the single chain
\citep{Dunkley:05}.  The first chain step within $\Delta \log
\mathcal{L} = 2$ of the best-fit parameters is taken as the start of the
converged sampling.  A Fisher-matrix approximation to the \pd\ fit is
used for the proposal density.  This code interpolates in log space
from a moderate number ($\sim \! \! 80$) of flux densities to
calcluate the \pd.  The other code is written in
$\mathrm{C}\!\!+\!\!+$, and is explicitly parallel.  It uses the
Gelman-Rubin criterion for burn-in \citep{Gelman:1992}, which is
based on computing the variance between chains and directly
provides the point of convergence.  This code does not use interpolation when
computing the \pd , but supports a more limited range of models.  The
proposal density is a multi-variate Gaussian estimated from the
previous fit steps, and is frozen in at burn-in.  We have checked
these codes against each other on simulated data, and find good
agreement.

Our \pd\ methodology is almost identical to that described in P09
except as follows. First, P09 explicitly fit for the mean of pixel
values in the map ($\mu$).  Since we can analytically predict the mean
of the \pd\ for a given set of model parameters, we simply shift the
mean to zero explicitly during construction.  The input map is also
mean-subtracted, and the uncertainty in this subtraction contributes
negligibly to our error budget.  Second, P09 fit to the instrument
noise explicity for each field except for the deepest section of their
map.  Instead, we marginalize over the noise for all fields in our
full fits, but use the measurements of \citet{Nguyen:2010} as a prior,
assuming an Gaussian uncertainty of 5\%.  At low flux densities, the
number of sources per beam is large, and hence the contribution to the
\pd\ is almost Gaussian.  Therefore, the values of $dN/dS$ for the
faintest flux densities probed and the noise level are nearly
degenerate, and hence fixing the noise will tend to under-estimate the
uncertainties in the model parameters at the faint end.

We have developed a simple simulation framework to test these codes
and their sensitivity to various effects such as $1/f$ noise.  As
inputs we consider two types of catalogs that should be representative
of the sub-mm sky: the P09 models, and the simulations of
\citet{fer08}.  The fits to the P09 models are easier to compare with
the inputs, but the \citet{fer08} models include clustering effects.

A fake sky is generated from the input catalogue, and scanned using
the pointing information from the actual SPIRE observations.
Different noise levels (white and $1/f$) can be specified.  These data
are then run through the same map-making pipeline as the real data.
In addition to the simulated science data, we also simulate
observations of Neptune using the same framework to determine the
beams we use when fitting the simulated data.  These simulations use
simple Gaussian beams with FWHMs similar to those measured on-orbit,
and account for characteristics of the data introduced by the mapping
pipeline, but do not simulate errors in the lower-level SPIRE pipeline
(pointing errors, crosstalk-corrections, etc.).  We use them to
quantify the effects of $1/f$ noise, uneven coverage, clustering, and
smoothing by the beam on our maps.  The SPIRE $1/f$ knee frequency is
a few mHz, corresponding to a spatial scale of approximately 3 degrees
for a scan speed of 30 arcseconds per second, and our map-making
algorithm reduces this already small amount as discussed in Levenson
et al.\ (2010, in prep).  We find that the remaining amount, as well
as the uneven coverage, introduces negligible bias in our fits, but
that clustering can have measureable effects on our largest maps, as
discussed in \S~\ref{subsec:filtering}.

In addition, we have determined the appropriate correction for
pixel-pixel correlations using the same framework and a large number
of simulated HerMES datasets.  We find that the correct normalization
factor varies with signal-to-noise ratio of the map, and whether it
has been additionally smoothed with the beam.  If the map is
beam-smoothed, then the beam area factor is approximately correct, if
slightly conservative for deeper fields; note that all of the maps in
P09 were beam-smoothed.  However, for deep, unsmoothed maps, this
procedure clearly overestimates the uncertainties (by about a factor
of 2 for GOODS-N).  Rather than derive individual correction factors
for each field, we have taken the more conservative approach of
finding the largest correction factor (which therefore increases the
uncertainties the most) for all of our fields, and applying it to all
un-smoothed data.  For the GOODS-N and Lockman-North observations, the
correct normalization factor (without smoothing) is less than $A_b/3$.
Because we do not have an exact formulation for this correction, we
conservatively adopt $2 A_b/5$.  For the smoothed observations, we
adopt the $A_b$ normalization, also conservatively; this means that
the two Lockman fields have the same correction factors, but the
(un-smoothed) GOODS-N field has a different one.

\subsection{Filtering}
\label{subsec:filtering}
\nobreak 
Clustering will affect the \pd\ distribution in two ways.  First, the
presence of clustering implies sample variance effects, so that the
SDP fields may not be representative of the all-sky number counts.
Second, the fact that the underlying counts are not Poisson
distributed would change the shape of the \pd\ distribution even if we
were somehow lucky enough to select a precisely average region of sky.
This effect can be modeled if all of the $n$-point statistics of the
source distribution are known \citep{barc92}.  The effect on the width
of the \pd\ is discussed in Appendix A of P09, although clustering is
not purely limited to changing the width of the distribution.  Only the
2-point function has been measured for the population sampled by
SPIRE, and even this is not known at the flux densities important for
our results.  The first issue is discussed in \S~\ref{subsec:systematics},
and the second here.  There are two effects: clustering on small
scales between individual SMGs, and clustering on larger scales between
groups of SMGs.

The framework for the clustering contribution to the \pd\ is given in
\citet{barc92,ti04}.  The contribution to the $n^{\mathrm{th}}$ moment
is proportional to $\int P_n\left( \mathbf{k} \right) \tilde{b}\left(
\mathbf{k} \right)^n \, d^2 \mathbf{k},$ where $P_n\left( \mathbf{k}
\right)$ is the power spectrum of the $n$-point correlation function,
and $\tilde{b}\left( \mathbf{k} \right)$ is the Fourier transform of
the beam.  $\tilde{b}$ falls rapidly with $\left| \mathbf{k} \right|$
(e.g., an 18 arcsecond FWHM Gaussian beam has a 1/e value at k = 1.2
arcmin$^{-1}$, and the higher powers fall off even more rapidly).
Thus, small scale clustering, which is implied by the measurements of,
e.g., \citet{Blain:2004}, is filtered out by the beam on scales of
less than about one to two arcminutes in our data.

Generically, $P_n$ falls rapidly with $\left| \mathbf{k} \right|$,
suggesting that high-pass filtering the maps may mitigate large-scale
clustering effects.  In particular, in the far-IR the power spectrum
of the two point correlation function ($P_2$) shows excess power above
Poissonian noise at scales larger than $10^{\prime}$
\citep{Lagache:2007, Viero:2009, Cooray:2010}. In order to reduce
ringing, our filter consists of a high-pass filter with a turn on at 6
arcmin convolved with a $\sigma = 1.8$ arcmin Gaussian.  Only the
Lockman-SWIRE field is large enough to be significantly affected
because the other fields are not much larger than this scale.  Since
the benefit of \pd\ analyses is at faint flux densities where most of
the CFIRB arises, and the shallow Lockman-SWIRE field has little
constraining power here, our main scientific results are minimally
affected by non-Poissonian clustering effects even if we ignore them.
In fact, we find that the differences between fits to simulated data
with and without clustering are well within the statistical errors
even without filtering.

\begin{figure}
\centering
\includegraphics[width=9cm]{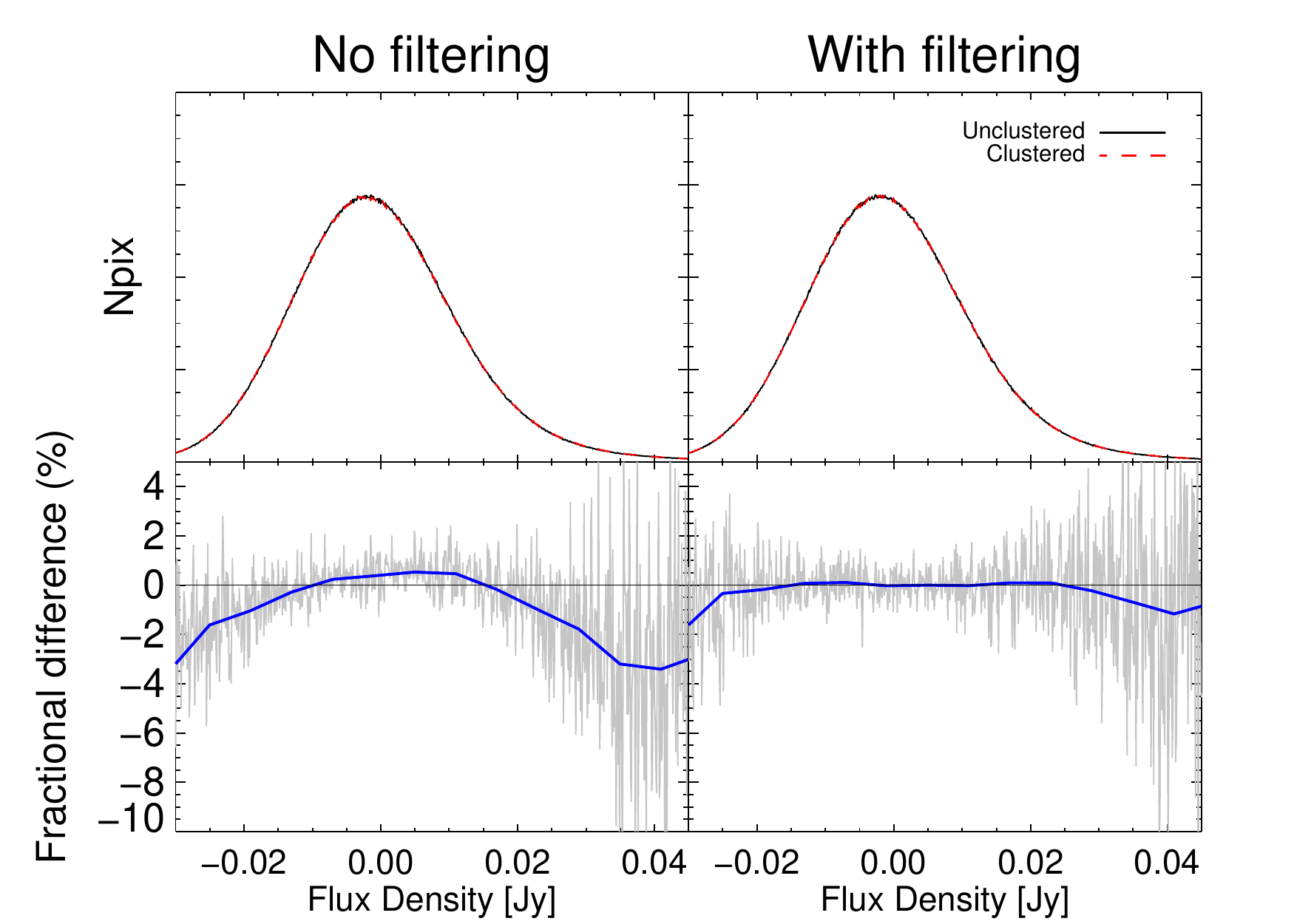}
\caption{ Comparison of the pixel histogram for simulated
  Lockman-SWIRE data with and without clustering, in the absence of
  high-pass filtering (left) and with a high-pass filter applied as
  described in the text (right).  The top panel shows the resulting
  pixel histograms, and the bottom shows the fractional difference
  between the clustered and unclustered pixel histograms.  The thick
  line in the bottom panel shows the average difference smoothed
  over a 3 mJy scale.  Without filtering, there is a clear trend in
  the fractional difference, but with filtering the difference is
  consistent with zero for the bulk of the pixel fluxes.  Even without
  filtering, the effect on the measured counts is smaller than the
  statistical errors. \label{fig:filterhisto} }
\end{figure}

Analysis of simulated data from \citet{fer08}, which has linear
clustering based on the assumption that infrared galaxies are tracers
of dark-matter fluctuations, shows that a high-pass filter is quite
effective at removing clustering signal for this data set.  We
construct two sets of simulated maps: one with clustering, and another
using the same catalogue but with clustering removed by randomizing
the source positions.  We then compare fits and pixel histograms for
both maps.  Because filtering will affect the \pd\ even in the absence
of clustering, comparing these to unclustered, unfiltered maps is not
useful.  The fits recover the input model accurately in both cases,
whereas if we do not filter $dN/dS$ is slightly underestimated at low
flux densities for the largest maps.  Smaller maps show no evidence
for bias.  A pixel histogram from such a simulation is shown in
figure~\ref{fig:filterhisto}. Such filtering is also effective at
removing infrared cirrus, although we have not tested this explicitly
in terms of the recovered fit parameters.  However, it is possible
that clustering signal on scales between one and six arcminutes could
affect our results.  This regime is currently not well characterized
and thus we could not model it quantitatively in our analysis.

\section{Model}
\label{sec:model}
\nobreak
The best approach for comparing a particular model to SPIRE data using
\pd\ is to generate pixel histograms as a function of the model
parameters and compare directly with our data.  However, not all
models have smoothly adjustable parameters, and furthermore, if the
model is a poor fit to the data this may provide little insight as to
at which flux densities the model disagrees with observations.  Hence,
we have followed P09 and fit simple, non-physical parametric models to
our data.  These models are defined by the values of the differential
number counts $dN/dS$ at a set of fixed flux densities (knots).
Observationally we can never do more than place a lower limit on the
total number of sources fainter than $S$, $N\left(< S\right)$ because
we can never measure all the way down to zero flux density, but  $dN/dS$
is better behaved because it only depends on the number of sources in
some small range.  The actual fit parameters are $\log_{10} dN/dS$ at
the knot positions.  The differential number counts must become
shallower than $S^{-2}$ at low flux densities or else the contribution
to the CFIRB diverges.  However, this turn-over may lie below the flux
densities probed by our data.  Therefore, in order to avoid biasing
our fits by excluding models which are too steep within the range of
our measurements (i.e., would overpredict the CFIRB if integrated
down to zero flux density), we assume that the number counts outside the largest
and smallest knot are zero; the problem of choosing the limits is
discussed below.

A \pd\ fit requires that the number counts model is continuous.
Therefore, we must choose a method of interpolating between the knots,
and for a finite number of knots, the interpretation of our results
depends on the interpolation method.  We consider two methods of
interpolation in this paper: first, as in P09, using power law
extrapolation between each knot (these are multiply-broken power-law fits),
and second, using a cubic spline in log-log space.  The first code
supports both methods, and the second only the former.  We do not
expect the fit parameters (i.e., $dN/dS$ at the knot positions) of
these models to be identical, since they have different meaning.

It is important to understand that the results of this paper are model
fits.  The fit results are not simply $dN/dS$ at the flux densities of
the knots, but instead are effectively integral constraints over some
region surrounding each knot.  Any excursion in the number counts that
lies entirely between two knots will affect at least both neighboring
knots, and likely others as well.  The flux density range that each
fit parameter is sensitive to depends on the interpolation scheme,
with the spline response more local to the knot.  Therefore, simply
reading off the values predicted by a theoretical or empirical number
counts model at the knot positions and comparing that with our fit
parameters is wrong since they are {\it integral constraints} over a
region surrounding the knot.  This is also true for more traditional
methods (i.e., simple number counts derived from individual galaxy
detections) because of the importance of the de-boosting corrections
for low signal-to-noise ratio detections.  A preferable approach is to
first find the best approximation to the differential counts of the
theoretical (or empirical) number counts model chosen for comparison
using either of our parametric models (for example, by fitting a
multiply-broken power law to the $dN/dS$ of the theoretical model
giving equal weighting to all fluxes, not just the values at the
knots) and comparing the parameters of that approximation to our
results.

The highest and lowest knot positions must be chosen with some care
because the differential number counts are assumed to be zero outside
this range.  Our criterion is based on examining the effects of
cutting the number counts at a given level on a selection of galaxy
evolution models from the literature.  We compare the predicted
\pd\ for each literature model truncated below a specified flux level
with the \pd\ without truncation, and find that our data are not
sensitive to a cut-off of less than 0.1 mJy at 250~\mum .  A similar
analysis shows that truncating the fit above 1 Jy is also
undetectable, with similar values for the other passbands.  From
simulations, we find that we can obtain good constraints if the second
knot lies approximately at the $1 \sigma$ instrumental noise.  Because
the number counts below our flux limit are unlikely to be well
described by a single knot all the way down to 0.1 mJy, the fit value
for this point should be treated with care; simulations indicate that
this is not a problem for the 2 mJy knot.  In order to avoid
over-tuning our fits to represent literature models, we adopt
approximately logarithmically spaced knots between these extremes.
The choice of the number of knots is somewhat arbitrary.  Neighboring
knots are very strongly correlated, and as the number increases the
correlations increase.  We have tried to chose the number of knots to
be as large as possible while keeping the correlations reasonably
small.

\section{Results}
\label{sec:results}
\nobreak
We fit all three fields simultanously, but each band independently.
The uncertainty in the instrumental noise is modeled as a single
multiplicative factor having a Gaussian prior with $\sigma = 5\%$.
Note that we are making the assumption that the timestream
instrumental noise is the same for all three fields as found in
\citet{Nguyen:2010}.  In addition to the SPIRE data, we also explore
the effects of including the FIRAS CFIRB prior \citep{Fixsen:1998} by
integrating $S dN/dS$ for our model down to the lowest knot and adding a term
to the likelihood that compares that value with the FIRAS measurement
and its error. This assumes that the CFIRB is entirely due to
discrete sources, and that flux densities outside the range of our
model contribute only negligibly.  We integrate the
\citet{Fixsen:1998} spectrum through the SPIRE passbands and adopt the
relative errors given in \citet{mar09}.  The uncertainty in the
relative calibration between FIRAS and SPIRE significantly affects the
utility of this prior.

\begin{table*}
 \begin{minipage}{90mm}
 \caption{Differential Number Count Constraints For a Multiply-Broken 
 Power-Law Model}
 \renewcommand{\arraystretch}{1.2}
 \begin{tabular}{@{}|rl|rl|rl@{}}
 \hline \hline
  \multicolumn{2}{c}{250 \mum } & \multicolumn{2}{c}{350 \mum } &
  \multicolumn{2}{c}{500 \mum } \\
 \hline
 Knot & $\log_{10} dN/dS$ & Knot & $\log_{10} dN/dS$ &
 Knot & $\log_{10} dN/dS$ \\
  $[\mathrm{mJy}]$ & $[\mathrm{deg}^{-2} \mathrm{Jy}^{-1}]$ &
  $[\mathrm{mJy}]$ & $[\mathrm{deg}^{-2} \mathrm{Jy}^{-1}]$ &
  $[\mathrm{mJy}]$ & $[\mathrm{deg}^{-2} \mathrm{Jy}^{-1}]$ \\
 \hline
  0.1 & $< 11.08 \left(1 \sigma\right)$ & 
     0.05 & $<11.20 \left(1 \sigma \right)$ &
     0.05 & $<11.28 \left(1 \sigma \right)$ \\
  2   & $7.05^{+0.33}_{-0.57} \pm 0.19$ & 
     2 & $6.94^{+0.13}_{-0.27} \pm 0.11$ & 
     2 & $6.82^{+0.11}_{-0.25} \pm 0.12$\\
  5   & $6.25^{+0.04}_{-0.13} \pm 0.05$ & 
     5 & $6.08^{+0.13}_{-0.25} \pm 0.08$ &
     5 & $5.65^{+0.19}_{-0.38} \pm 0.09$ \\
 10 & $5.919^{+0.028}_{-0.063} \pm 0.011$ & 
    10 & $5.78^{+0.05}_{-0.11} \pm 0.04$ &
    10 & $5.39^{+0.09}_{-0.18} \pm 0.03$ \\
 20 & $5.139^{+0.013}_{-0.035} \pm 0.025$ & 
    20 & $4.976^{+0.026}_{-0.061} \pm 0.024$ & 
    20 & $4.57^{+0.05}_{-0.12} \pm 0.03$ \\
 45   & $4.038^{+0.015}_{-0.033} \pm 0.031$ & 
    45 & $3.742^{+0.026}_{-0.061} \pm 0.051$ & 
    45 & $2.91^{+0.07}_{-0.16} \pm 0.04$ \\
 100  & $2.596^{+0.025}_{-0.058} \pm 0.044$ & 
   100 & $1.80^{+0.07}_{-0.16} \pm 0.10$ &
   100 & $0.96^{+0.22}_{-0.38} \pm 0.06$ \\
 200  & $1.42^{+0.05}_{-0.14} \pm 0.08$ & 
   200 & $0.87^{+0.14}_{-0.28} \pm 0.08$ &
   200 & $0.00^{+0.51}_{-0.92} \pm 0.07$ \\
 450  & $0.57^{+0.13}_{-0.24} \pm 0.26$ & 
   750 & $-0.65^{+0.39}_{-0.78} \pm 0.30$ &
   600 & $-1.43^{+0.96}_{-2.09} \pm 0.29$ \\
 1000 & $-0.45^{+0.31}_{-0.60} \pm 0.20$ & & \\
 \hline
 \end{tabular}
  Marginalized fit parameters for a multiply-broken power-law model
  from a joint analysis of all three fields
  without using the FIRAS CFIRB prior.  The quoted uncertainties
  are the $68.3\%$ confidence intervals for the statistical error followed
  by the estimated systematic uncertainty, except for the first knot where
  the $1 \sigma$ upper limit is given.   \label{tbl:brokpownoprior}
 \end{minipage}
\end{table*}

\begin{table}
 \caption{Differential Number Counts Constraints For a Multiply-Broken
   Power-Law Model with the FIRAS prior}
 \renewcommand{\arraystretch}{1.2}
 \begin{tabular}{@{}|rl|rl|rl@{}}
 \hline \hline
  \multicolumn{2}{c}{250 \mum } & \multicolumn{2}{c}{350 \mum } &
  \multicolumn{2}{c}{500 \mum } \\
 \hline
 Knot & $\log_{10} dN/dS$ & Knot & $\log_{10} dN/dS$ &
 Knot & $\log_{10} dN/dS$ \\
  $[\mathrm{mJy}]$ & $[\mathrm{deg}^{-2} \mathrm{Jy}^{-1}]$ &
  $[\mathrm{mJy}]$ & $[\mathrm{deg}^{-2} \mathrm{Jy}^{-1}]$ &
  $[\mathrm{mJy}]$ & $[\mathrm{deg}^{-2} \mathrm{Jy}^{-1}]$ \\
 \hline
  0.1 & $< 9.38 \left(1 \sigma\right)$ & 
     0.05 & $<5.42 \left(1 \sigma \right)$ &
     0.05 & $<4.37 \left(1 \sigma \right)$ \\
  2   & $7.39^{+0.06}_{-0.17}$ & 
     2 & $7.87^{+0.08}_{-0.19}$ & 
     2 & $7.70^{+0.09}_{-0.21}$ \\
  5   & $5.83^{+0.15}_{-0.25}$ & 
     5 & $5.24^{+0.32}_{-0.56}$ &
     5 & $5.50^{+0.29}_{-0.52}$ \\
 10 & $5.978^{+0.027}_{-0.071}$ & 
    10 & $5.87^{+0.05}_{-0.10}$ &
    10 & $5.13^{+0.10}_{-0.21}$ \\
 20 & $5.13^{+0.02}_{-0.04}$ & 
    20 & $4.960^{+0.028}_{-0.067}$ & 
    20 & $4.686^{+0.039}_{-0.094}$ \\
 45   & $4.041^{+0.014}_{-0.034}$ & 
    45 & $3.744^{+0.026}_{-0.062}$ & 
    45 & $2.82^{+0.07}_{-0.15}$ \\
 100  & $2.591^{+0.027}_{-0.062}$ & 
   100 & $1.82^{+0.07}_{-0.16}$ &
   100 & $1.10^{+0.19}_{-0.35}$ \\
 200  & $1.42^{+0.06}_{-0.14}$ & 
   200 & $0.81^{+0.16}_{-0.29}$ &
   200 & $-0.08^{+0.54}_{-0.97}$ \\
 450  & $0.58^{+0.12}_{-0.22}$ & 
   750 & $-0.69^{+0.16}_{-0.29}$ &
   600 & $-1.45^{+0.99}_{-2.01}$ \\ 
 1000 & $-0.44^{+0.32}_{-0.62}$ & & \\
 \hline
 \end{tabular}
  Marginalized fit parameters for a multiply-broken power-law model
  from a joint analysis of all three fields including the
  FIRAS CFIRB prior of \citet{Fixsen:1998}.  Quoted uncertainties are
  $68.3\%$ confidence intervals, except for the first knot where
  $1 \sigma$ upper limits are given.  The systematic uncertainties
  are the same as in table~\ref{tbl:brokpownoprior}.
   \label{tbl:brokpowprior}
\end{table}

\begin{figure*}
\begin{minipage}{130mm}
 \centering
 \includegraphics[width=14cm]{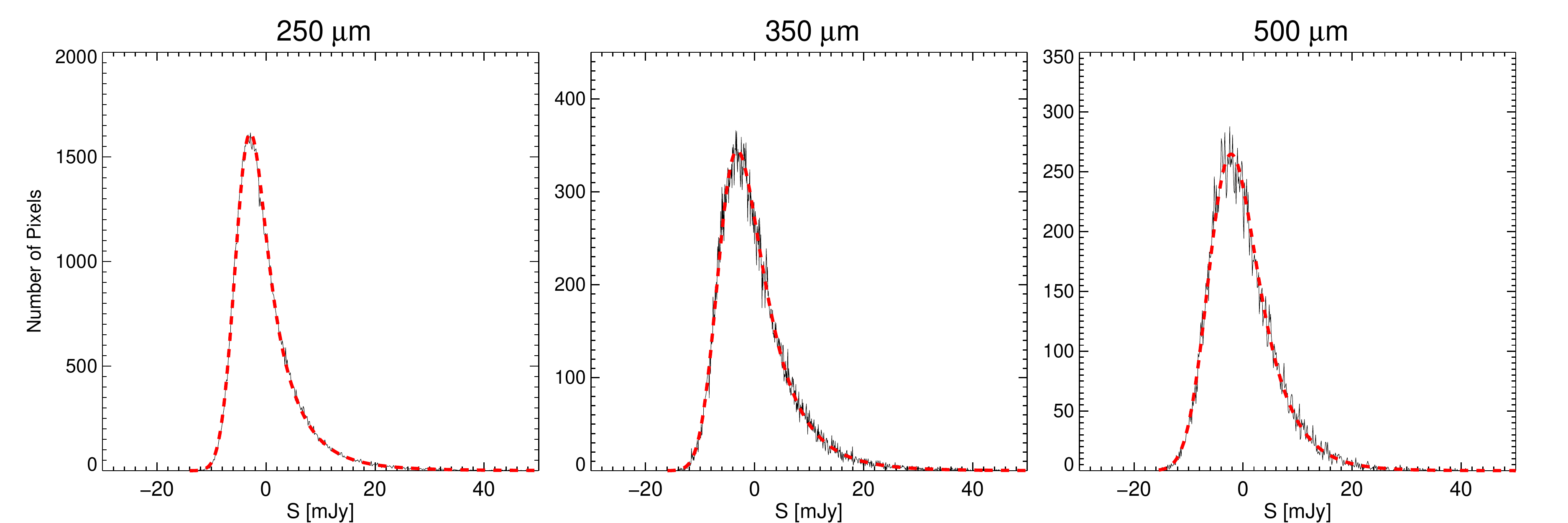}
 \caption{Comparison of the GOODS-N pixel histograms (solid lines)
  to the best fit model to all three fields (dashed lines) using
  the multiply-broken power law fit and not including the FIRAS
  prior. \label{fig:fitplot} }
\end{minipage}
\end{figure*}

\begin{figure*}
\begin{minipage}{130mm}
 \centering
 \includegraphics[width=14cm]{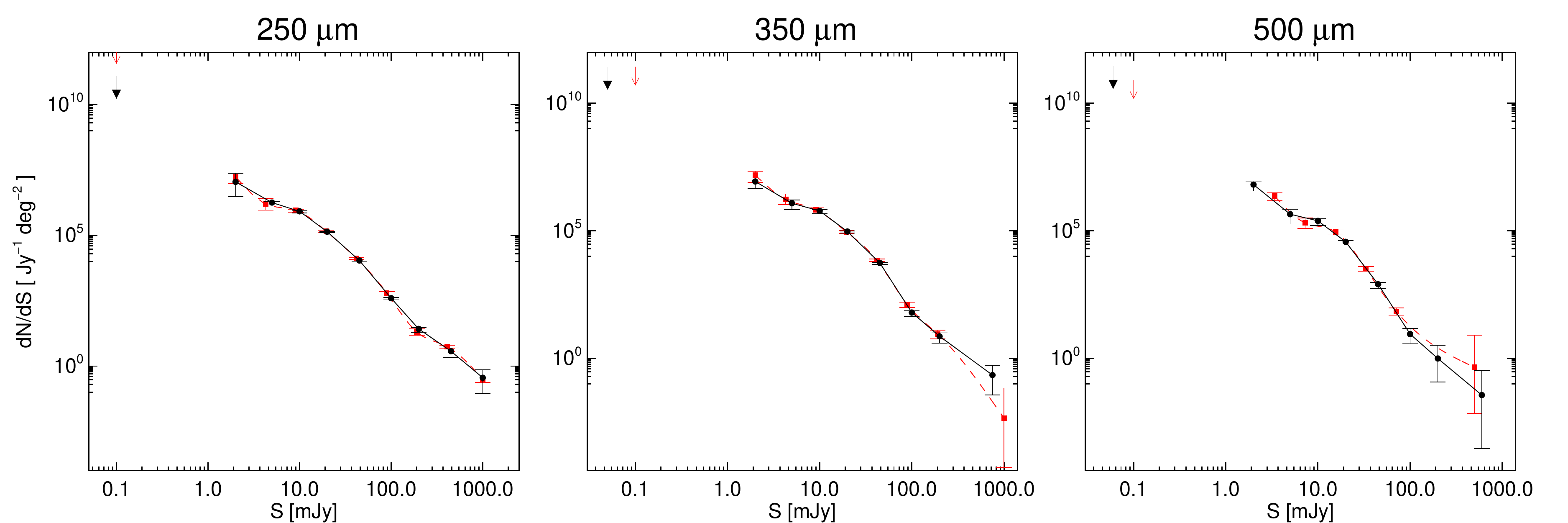}
 \caption{Comparison of the multiply-broken power-law (solid lines) and
   spline (dashed lines) \pd\ fits for the differential number
   counts to the three SDP fields simultaneously, without the FIRAS
   prior; $1\sigma$ upper limits are shown as arrows.  For this
   comparison, only statistical errors are shown.
  \label{fig:comparemodel} }
\end{minipage}
\end{figure*}

The best fit multiply broken power-law fit is compared with the
GOODS-N data in figure~\ref{fig:fitplot}, and the parameters are given
in tables~\ref{tbl:brokpownoprior} and \ref{tbl:brokpowprior}, and for
the spline interpolation fits in tables~\ref{tbl:splinenoprior} and
\ref{tbl:splineprior}.  The correlations between adjacent knots are
large and negative\footnote{Covariance matrices are provided at
  http://www.hermes.sussex.ac.uk/}, with typical correlation
coefficients of $-0.5$ to $-0.8$.  The two models are compared with
each other in figure~\ref{fig:comparemodel}. The two interpolating
models (spline and multiply-broken power-law) produce very similar
results.  As discussed previously, these are model fits, not
independent number counts, and since the parameterizations differ,
directly comparing the values at the knot positions is not entirely
correct.  Nonetheless, the agreement is clear.  Also, because the
models were fit to the same data their results should not be coadded:
they are both presented to demonstrate that similar results are
obtained with using independent codes.

Since the agreement is so good, we express no preference of one model
versus the other (multiply-broken power-law versus spline).  However,
we note that the spline model has a narrower flux density window
function about each knot, and thus represents the differential number
counts of the knot position slightly more accurately locally than the
power-law model.  For comparison to other number counts models, one
can either (i) select the fits with the FIRAS prior, which assumes
that the remaining portion of the CFIRB unaccounted for by our
priorless fits is encompassed in the range between the upper limit on
$dN/dS$ at 0.1 mJy and the 2 mJy knot (at 250 \mum) -- this method is
simpler; or (ii) select the fits without the CFIRB prior, in which
case the prior should be applied independently.  The latter does not
require that the model share the same assumptions about the number
counts at low flux densities as our fits.

\begin{figure*}
\begin{minipage}{130mm}
 \centering
 \includegraphics[width=14cm]{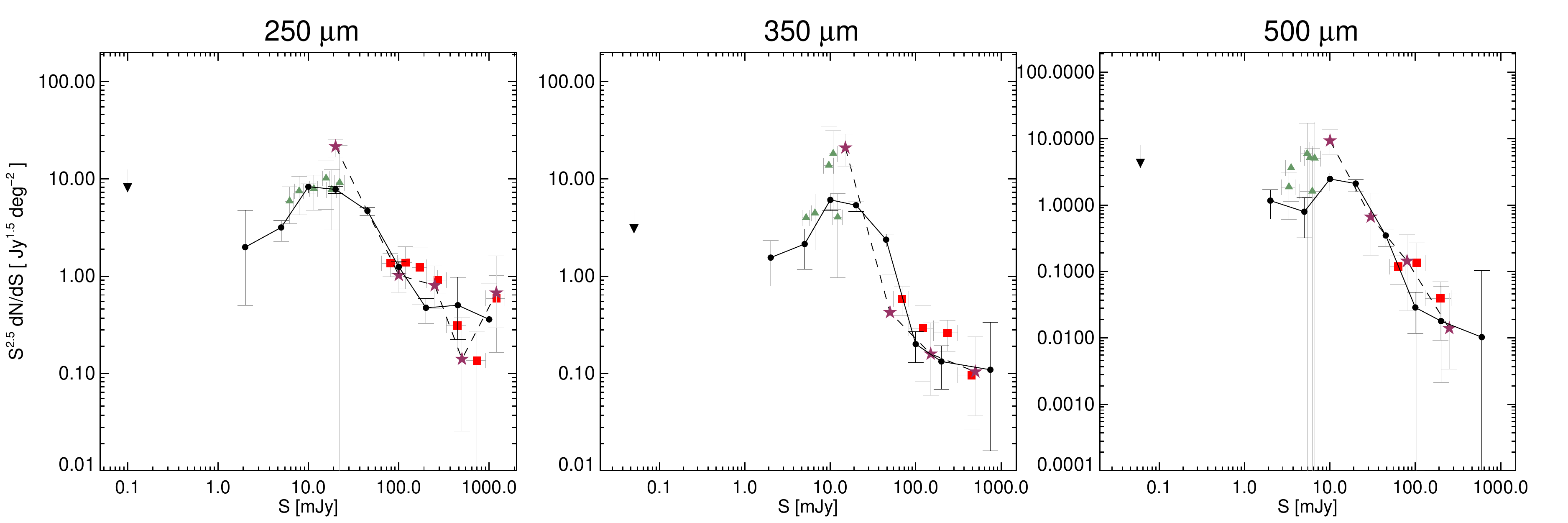}
 \caption{Comparison of Euclidean-normalized multiply-broken power law
   differential number counts fits (solid lines and circles)
   with previous balloon-based measurements from BLAST, not using the
   FIRAS prior.  The BLAST \pd\ analysis (P09) is shown as dashed
   lines and stars, the stacking analysis of
   \citet{Bethermin:2010b} as triangles and the source
   extraction analysis from the same reference as squares. Here
   the combined statistical and systematic errors are shown.
\label{fig:brokpow} }
\end{minipage}
\end{figure*}

\begin{figure*}
\begin{minipage}{130mm}
 \centering
 \includegraphics[width=14cm]{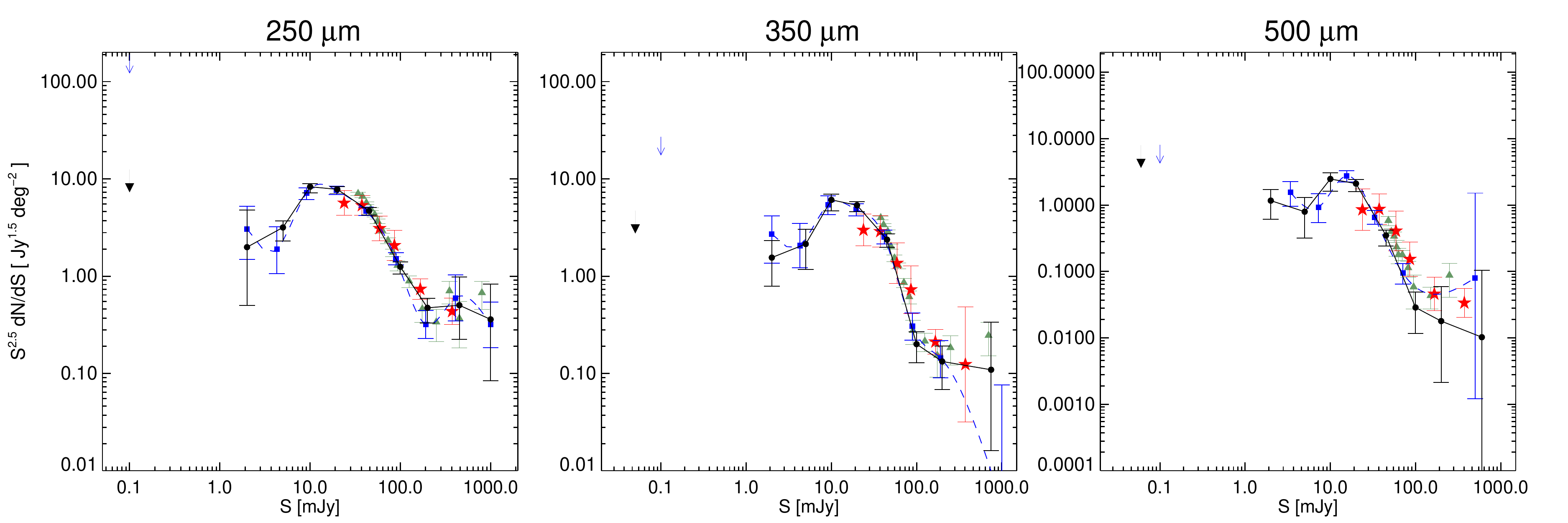}
 \caption{Comparison of Euclidean normalized SPIRE \pd\ differential
   number counts (solid lines/circles and dashed
   lines/squares for the multiply-broken power law and spline models,
   respectively, not using the FIRAS prior in both cases) with other
   SPIRE number counts: first, an analysis of the same data using
   source extraction techniques \citep[][red stars]{Oliver:2010b}, and
   second the H-ATLAS source extraction on an independent field
   \citep[][triangles]{Clements:2010}. The errors are the
   combined statistical and systematic
   errors. \label{fig:brokpowoliver} }
\end{minipage}
\end{figure*}

\begin{figure}
\centering
\includegraphics[width=9cm]{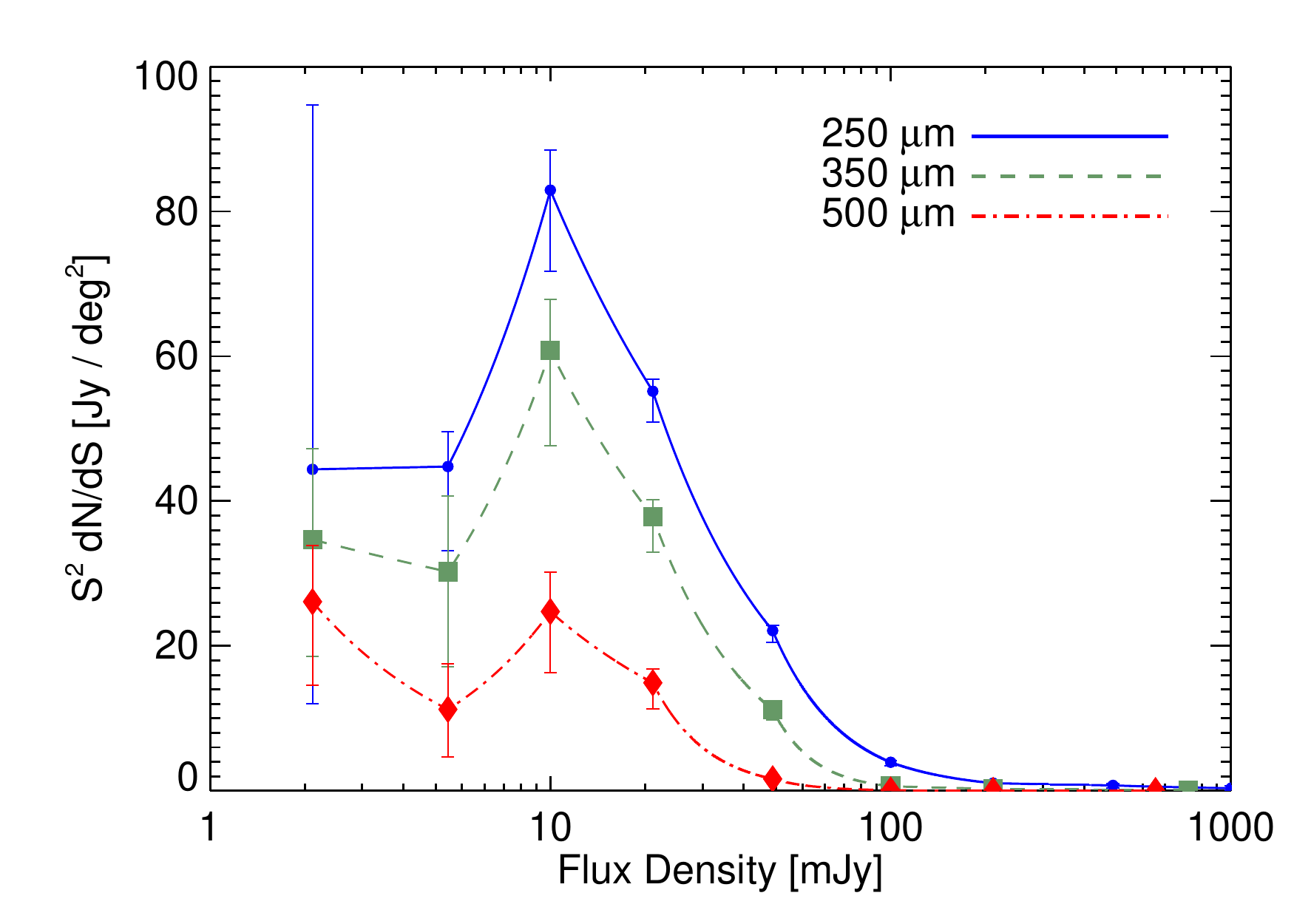}
\caption{The contribution to the cosmic far-infrared background from
  each flux range for the multiply-broken power-law model versus the
  log of the flux density.  The results for the spline interpolation
  model are almost identical.  The integral of the curve over the
  $\log$ flux density is proportional to the total flux contribution in each
  band. \label{fig:cfirb_contrib}}
\end{figure}

Our fits are compared with other measurements in
figures~\ref{fig:brokpow} and \ref{fig:brokpowoliver}.  Ignoring the
lowest knot (where only an upper limit is available), our fits predict
a CFIRB flux density of $0.54 \pm 0.08$, $0.39 \pm 0.06,$ and $0.16
\pm 0.03\, \mathrm{MJy}\, \mathrm{sr}^{-1}$ from all sources down to 2
mJy in the three bands; the dominant error in all cases is due to the
15 per cent calibration uncertainty of SPIRE.  The contribution from
each flux range is shown in figure~\ref{fig:cfirb_contrib}.  The CFIRB
from \citet{Fixsen:1998} integrated over the SPIRE bands is $0.85 \pm
0.19$, $0.65 \pm 0.19$, and $0.39 \pm 0.10$ MJy sr$^{-1}$,
respectively, so our fits therefore account for $64 \pm 16, 60 \pm 20$
and $43 \pm 12$ percent in the SPIRE 250, 350, and 500~\mum\ bands,
respectively.  We expect to resolve a smaller fraction of the CFIRB at
longer wavelengths because the size of the SPIRE beam is proportional
to wavelength, and hence the 500~\mum\ band is more confused.  Here
the errors are dominated by the uncertainty in the FIRAS measurement.
We find marginalized values for the instrumental noise that are 1.02,
1.1, and 1.01 times the values given in table~\ref{tbl:data} at 250,
350, and 500~\mum , respectively, giving a $\chi^2$ of 4.2 for 3
degrees of freedom.  Hence, our instrumental noise values are
consistent with the \citet{Nguyen:2010} prior.

\begin{table*}
 \begin{minipage}{90mm}
 \caption{Differential Number Counts Constraints For a Spline Model}
 \renewcommand{\arraystretch}{1.2}
 \begin{tabular}{@{}|rl|rl|rl@{}}
 \hline \hline
  \multicolumn{2}{c}{250 \mum } & \multicolumn{2}{c}{350 \mum } &
  \multicolumn{2}{c}{500 \mum } \\
 \hline
 Knot & $\log_{10} dN/dS$ & Knot & $\log_{10} dN/dS$ &
 Knot & $\log_{10} dN/dS$ \\
  $[\mathrm{mJy}]$ & $[\mathrm{deg}^{-2} \mathrm{Jy}^{-1}]$ &
  $[\mathrm{mJy}]$ & $[\mathrm{deg}^{-2} \mathrm{Jy}^{-1}]$ &
  $[\mathrm{mJy}]$ & $[\mathrm{deg}^{-2} \mathrm{Jy}^{-1}]$ \\
 \hline
  0.1 & $< 10.29 \left(1 \sigma\right)$ & 
     0.05 & $<11.43 \left(1 \sigma \right)$ &
     0.05 & $<10.91 \left(1 \sigma \right)$ \\
  2   & $7.26^{+0.10}_{-0.17} \pm 0.19$ & 
     2 & $7.18^{+0.15}_{-0.28} \pm 0.11$ & 
     3.4 & $6.36^{+0.13}_{-0.18} \pm 0.10$ \\
  4.3   & $6.54^{+0.10}_{-0.12} \pm 0.06$ & 
     4.3 & $6.24^{+0.21}_{-0.21} \pm 0.09$ &
     7.3 & $5.31^{+0.19}_{-0.21} \pm 0.05$ \\
 9.1 & $5.837^{+0.059}_{-0.056} \pm 0.013$ & 
    9.1 & $5.831^{+0.081}_{-0.090} \pm 0.042$ &
    15.5 & $4.961^{+0.074}_{-0.085} \pm 0.029$ \\
 19.5 & $5.230^{+0.029}_{-0.032} \pm 0.024$ & 
    19.5 & $4.959^{+0.053}_{-0.061} \pm 0.025$ & 
    33.2 & $3.511^{+0.094}_{-0.095} \pm 0.034$ \\
 41.8   & $4.036^{+0.032}_{-0.036} \pm 0.030$ & 
    41.8 & $3.849^{+0.051}_{-0.050} \pm 0.048$ & 
    71 & $1.85^{+0.14}_{-0.16} \pm 0.063$ \\
 89.3  & $2.802^{+0.045}_{-0.050} \pm 0.040$ & 
   89.3 & $2.11^{+0.10}_{-0.11} \pm 0.076$ &
   500 & $-0.64^{+1.25}_{-1.80} \pm 0.028$ \\
 191  & $0.20^{+0.24}_{-0.34} \pm 0.080$ & 
   191 & $0.96^{+0.16}_{-0.19} \pm 0.075$ & \\
 408  & $1.002^{+0.064}_{-0.068} \pm 0.26$ & 
   1000 & $-2.34^{+1.82}_{-1.92} \pm 0.31$ & \\
 1000 & $0.18^{+0.09}_{-0.10} \pm 0.20$ & & \\
 \hline
 \end{tabular}
  Marginalized fit parameters for a spline model
  from a joint analysis of all three fields.  Quoted
  uncertainties are as in table~\ref{tbl:brokpownoprior}.
  \label{tbl:splinenoprior}
\end{minipage}
\end{table*}

\begin{table}
 \caption{Differential Number Counts Constraints For a Spline Model with the
   FIRAS prior}
 \renewcommand{\arraystretch}{1.2}
 \begin{tabular}{@{}|rl|rl|rl@{}}
 \hline \hline
  \multicolumn{2}{c}{250 \mum } & \multicolumn{2}{c}{350 \mum } &
  \multicolumn{2}{c}{500 \mum } \\
 \hline
 Knot & $\log_{10} dN/dS$ & Knot & $\log_{10} dN/dS$ &
 Knot & $\log_{10} dN/dS$ \\
  $[\mathrm{mJy}]$ & $[\mathrm{deg}^{-2} \mathrm{Jy}^{-1}]$ &
  $[\mathrm{mJy}]$ & $[\mathrm{deg}^{-2} \mathrm{Jy}^{-1}]$ &
  $[\mathrm{mJy}]$ & $[\mathrm{deg}^{-2} \mathrm{Jy}^{-1}]$ \\
 \hline
  0.1 & $< 8.743 \left(1 \sigma\right)$ & 
     0.05 & $<7.63 \left(1 \sigma \right)$ &
     0.05 & $<6.81 \left(1 \sigma \right)$ \\
  2   & $7.281^{+0.067}_{-0.081}$ & 
     2 & $7.335^{+0.066}_{-0.077}$ & 
     3.4 & $6.578^{+0.064}_{-0.083}$ \\
  4.3   & $6.565^{+0.082}_{-0.094}$ & 
     4.3 & $6.26^{+0.23}_{-0.26}$ &
     7.3 & $5.37^{+0.14}_{-0.22}$ \\
 9.1 & $5.817^{+0.049}_{-0.052}$ & 
    9.1 & $5.78^{+0.10}_{-0.11}$ &
    15.5 & $4.91^{+0.090}_{-0.093}$ \\
 19.5 & $5.241^{+0.027}_{-0.028}$ & 
    19.5 & $4.983^{+0.058}_{-0.063}$ & 
    33.2 & $3.55^{+0.09}_{-0.10}$ \\
 41.8   & $4.023^{+0.033}_{-0.031}$ & 
    41.8 & $3.831^{+0.054}_{-0.055}$ & 
    71 & $1.82^{+0.15}_{-0.16}$ \\
 89.3  & $2.786^{+0.045}_{-0.049}$ & 
   89.3 & $2.13^{+0.10}_{-0.11}$ &
   500 & $-0.63^{+1.11}_{-1.80}$ \\
 191  & $-0.08^{+0.26}_{-0.32}$ & 
   191 & $0.95^{+0.16}_{-0.18}$ & \\
 408  & $0.957^{+0.065}_{-0.075}$ & 
   1000 & $-2.12^{+1.06}_{-1.83}$ & \\
 1000 & $0.186^{+0.084}_{-0.088}$ & & \\
 \hline
 \end{tabular}
  Marginalized fit parameters for a spline interpolation model from a
  joint analysis of all three fields including the FIRAS CFIRB prior
  of \citet{Fixsen:1998}.  The systematic uncertainties are the same
  as in table~\ref{tbl:splinenoprior}.
   \label{tbl:splineprior}
\end{table}

\subsection{Systematic Effects}
\label{subsec:systematics}
\nobreak
Our basic tool for estimating the importance of a particular
systematic is to compute the $\Delta \log \mathcal{L}\,$ between the
\pd\ with and without the effect for maps the same size and depth as
our data.  We use the P09 best fit model as a basis for this
computation.  Recall that a $\Delta \log \mathcal{L}\,$ of 0.5
corresponds roughly to a $1 \sigma$ statistical error.

Because different parts of the map are sampled by different
bolometers, and the beam shape varies across the bolometer array, the effective
beam will vary over the map.  We evaluated this effect by choosing 200
random pixels in our maps and computing the fractional contribution of
each bolometer to each pixel.  We then built per-bolometer maps from
our Neptune observations, and combined these to find the effective
beam at each of these locations.  The beam varies across the map in a
complicated fashion because even in our deepest map each pixel only
samples a limited subset of bolometers.  This produces
significant variation in the \pd\ with position.  In general, the
\pd\ computed for the bolometer-averaged beam does not have to be the
same as the \pd\ computed for each bolometer and then averaged across the
array (which is the \pd\ of the entire map).  To evaluate the
importance of this variation, we compare the \pd\ for the average beam
to the \pd\ computed for all 200 pixels and then averaged.  The
$\Delta \log \mathcal{L}\,$ of this comparison is $< 0.01$, so for our
analysis this is negligible.

Although we masked each map to exclude the low-coverage edges, the
HerMES SDP observations were not dithered between repeats so there are
significant variations in the number of measurements per pixel even
within the high-coverage regions ($\sim 20\%$).  This will introduce
slight non-Gaussian tails to the instrument noise distribution.  We
simulated this effect in two ways. First, we generated random
realizations of the instrument noise, including the uneven coverage,
and compared the \pd\ using the resulting noise to the \pd\ assuming
the noise is purely described by the average $\sigma$, and found
negligible $\Delta \log \mathcal{L}\,$.  Second, our end-to-end
simulations (also including $1/f$ noise) implicitly include uneven
coverage effects, and we found no bias in the recovered parameters.
Future HerMES observations will include dithering, which will also
have the benefit of improving the homogeneity of the beam.

\citet{Nguyen:2010} explore the noise characteristics of
the SDP maps by carrying out `Jack-Knife' tests on the data.  Their
findings are generally consistent with the expected noise properties,
but it is difficult to rule out some additional level of non-Gaussian
noise beyond the $1/f$ behavior we have simulated.  Directly computing
the effects of the Jack-Knife noise histograms on our model shows that
any additional non-Gaussianity has negligible effects for the SDP
data, down to our lowest constrained knot (2 mJy).  This may not be
true for future observations where the larger field sizes will reduce
the statistical error considerably.

To test the sensitivity to the beam model, we use an alternative set
of Neptune observations with a much smaller number of repeats and
coarser sampling.  Furthermore, the pointing of these observations is
not corrected for the small offset between the {\it Herschel} and
SPIRE clocks, and hence they suffer from pointing drift relative to
the maps of the science fields\footnote{The SPIRE clock speed differs
  by a very small amount from the Herschel clock, resulting in a
  cumulative pointing drift with time in SPIRE maps.  The magnitude of
  the effect is 0.7 arcseconds per hour, with rephasing occuring when
  ``PCAL" internal calibrations are made.  }.  The pipeline nominally
corrects for this offset.  However, to be conservative, we allow for
the possiblity that the pointing drift might not affect the beam maps
in the same way as the science maps: we repeat the fits using the
alternative beam, and use the difference in the results as an estimate
of the beam systematic.  This is the dominant identified systematic
effect, with $\Delta \mathcal{L} \sim 0.3$.  We take the differences
between the recovered parameters between the two beams as the
systematic error on each knot as given in
tables~\ref{tbl:brokpownoprior} and \ref{tbl:splinenoprior}.  As our
understanding of the SPIRE beams improves, it should be possible to
decrease this error.

To further explore issues of pointing drift, we have constructed a
simple drift model for the GOODS-N field using Jack-Knife comparisons.
We then generate simulated maps with and without applying this model.
Because the effect of the model is largely to twist successive
observations relative to each other, this has very little effect on
the \pd , amounting to $\sim 0.1 \sigma$ relative to the statistical
errors.

\begin{figure}
\centering
\includegraphics[width=9cm]{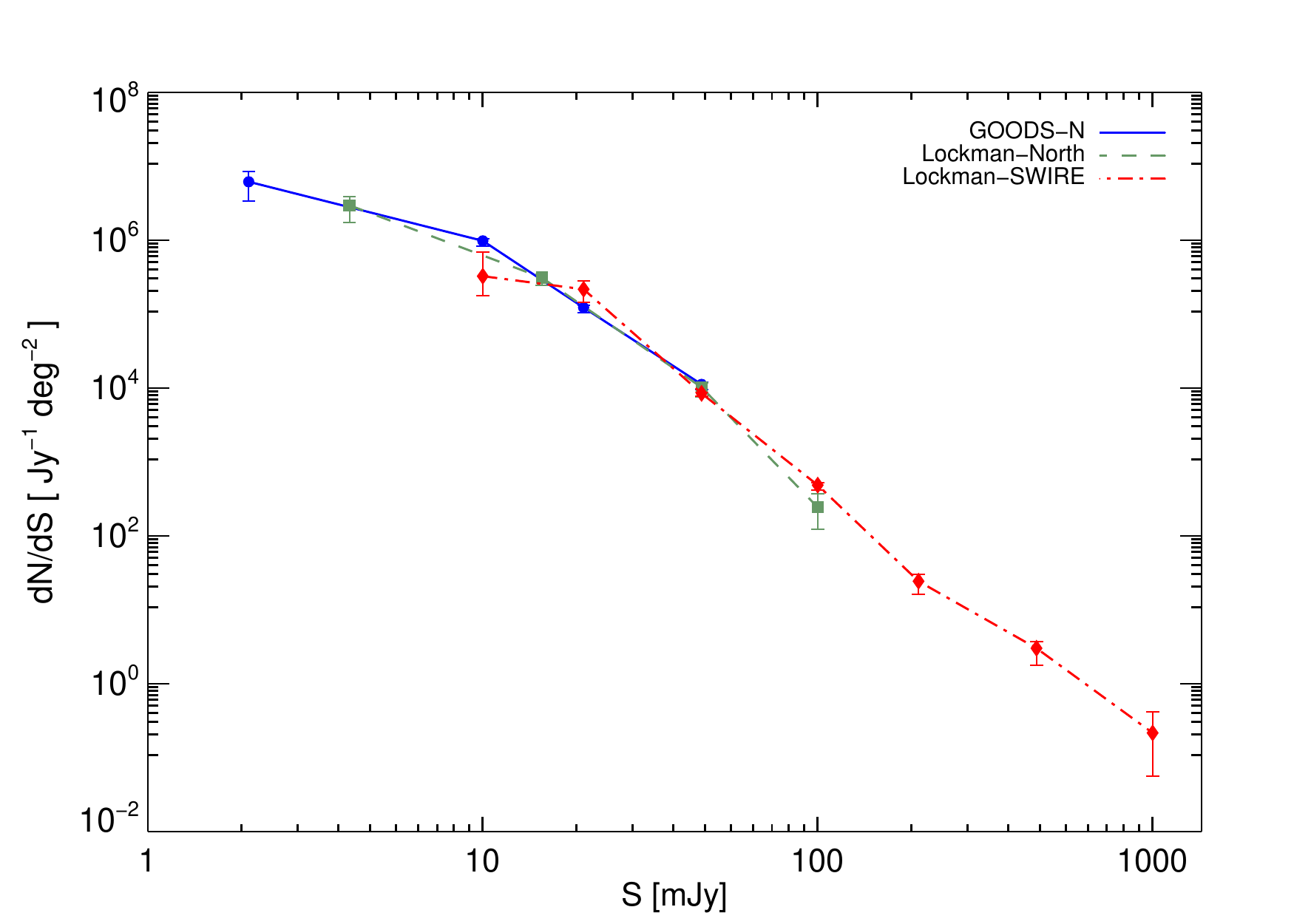}
\caption{Comparison of 250~\mum\ differential number counts derived
  from the three fields for the multiply-broken power-law model. Only
  statistical errors are shown and the FIRAS prior is not included.
 \label{fig:fieldcompare} }
\end{figure}

While our results should represent the number counts in our fields
quite well, sample variance means that they may not perfectly
represent the number counts we would obtain with an infinitely large
field.  If we make the strong assumptions that the SPIRE clustering
properties measured in \citet{Cooray:2010} apply equally at all flux
densities (and, in particular, to depths 10 times greater than they
were measured), that the redshift distribution of our sources is
independent of measured brightness, and that the source population
peaks at $z=1.5$, then a simple analytic computation suggests that
sample variance could be a 20 per cent effect on the total number
counts in GOODS-N.  More empirically, if we split the Lockman-SWIRE
field into sub-regions there is evidence for variation in the pixel
histogram from sub-field to sub-field.  However, after applying our
high-pass filter the variations are no longer statistically
signficant; that is, the difference between sub-fields lies in the mean
rather than the shape of the histograms.  Alternatively, mean
subtracting each sub-field produces the same effect.  This suggests
that the fact that our measurements are not sensitive to the mean map
values (and therefore are mean-subtracted) may provide some protection
against sample variance effects.  Since estimating the size of this
effect is highly model dependent, we have not attempted to include
this in our error budget.  We do, however, fit the three fields
independently at 250~\mum\ and compare the results, although the
different depths complicate this somewhat.  For simplicity, we do not
marginalize over the instrument noise in this fit, since the only
purpose is to compare the three fields.  This is shown in
figure~\ref{fig:fieldcompare}.  Within the uncertainties, the fits are
consistent.  The post-SDP HerMES observations will allow sample
variance effects to be better quantified.

We do not include the effects of the SPIRE calibration uncertainty
($\sim 15\%$ across all bands) in our error budget, except when using
the CFIRB prior or computing the fraction of the FIRAS measurement
accounted for by our model.  However, any updates to the SPIRE
calibration are easily incorporated into our results without
refitting: if the flux scale is multiplied by a factor $\alpha$ the
knot positions $K_i \mapsto \alpha K_i$ and the knot values decrease
by $- \log_{10} \alpha$.

\section{Discussion}
\label{sec:discussion}
\nobreak 
In general, our results agree well with those of P09, except for the
faintest fluxes fully constrained by their analysis.  For example, at
250~\mum\ they find $\log_{10} dN/dS = 5.58^{+0.07}_{-0.11}$ at 20
mJy, while our result is $5.139^{+0.012}_{-0.033} \pm 0.025$.
However, as discussed earlier, they did not marginalize over the
instrumental noise for their deepest field, so their errors may be
somewhat underestimated here.  There is also some evidence from
simulated data that the small number of knots and knot placement right
at the break in the number counts may have biased this knot in the P09
analysis.  We find good agreement with the stacking analysis of the
BLAST data, but see some mild disagreement at higher fluxes for direct
counts of the same data \citep{Bethermin:2010b} as shown in
figure~\ref{fig:brokpow}.

The deepest number counts available at these wavelengths are the
result of a semi-traditional source extraction method on the same
HerMES data set \citep{Oliver:2010b}.  These are compared in
figure~\ref{fig:brokpowoliver}.  Where there is overlap, the agreement
is good.  A similar analysis was carried out using H-ATLAS SDP data
by \citet{Clements:2010}; this is also shown.  Unlike 
the HerMES source extraction and \pd\ analysis, the H-ATLAS counts
are 250~\mum-selected at all wavelengths, and hence may not entirely
probe the same point source population.  Nonetheless, again the
agreement with the HerMES results is good.

A few features are worth noting.  First, we clearly detect a break in
the number counts around at 20 mJy in all bands at high significance.
However, the SPIRE data alone do not detect the change in slope in
$dN/dS$ necessary to keep the CFIRB finite, as the differential counts
continue to rise to the lowest limit of our analysis more steeply than
$S^{-2}$.  When the FIRAS prior is added, a break is present, but this
mostly affects the lowest flux knot, for which we can only provide an
upper limit (the effects on the other knots are mostly due to the
strong correlations between knots; the FIRAS prior changes the
structure of the correlations significantly at low flux densities).
Second, there is possible weak ($\sim 1\sigma$) evidence for a `bump'
in the differential counts around 400 mJy at 250~\mum .  There is no
evidence at 350 and 500~\mum\ around this flux density, but the error
bars are large.  However, this bump is present in the independent
H-ATLAS field \citep{Clements:2010} at 250~\mum .  The cause is
unclear; lensing is an intriguing possibility, but we would expect
the signature of lensing to be larger at 500 \mum\ due to
stronger negative $K$-correction effects \citep[e.g.][]{Negrello:2007}.

We can compare our model fits to the confusion noise estimates of
\citet{Nguyen:2010} measured using a different technique.  The
often-used criterion of one source per every $n$ beams is difficult to
use, since its translation into the effects on observations depends
strongly on the underlying model \citep{ti04}.  Hence, we adopt the
square root of the variance of the source contribution to the pixel
distribution ($\sigma_{\mathrm{conf}}$) as our measure.  We find
values of $6.5/6.4/6.1 \pm 0.2$ mJy in the three bands, slightly
higher than the \citet{Nguyen:2010} values, but by less than
$2 \sigma$.

\begin{figure*}
\begin{minipage}{130mm}
\centering
\includegraphics[width=14cm]{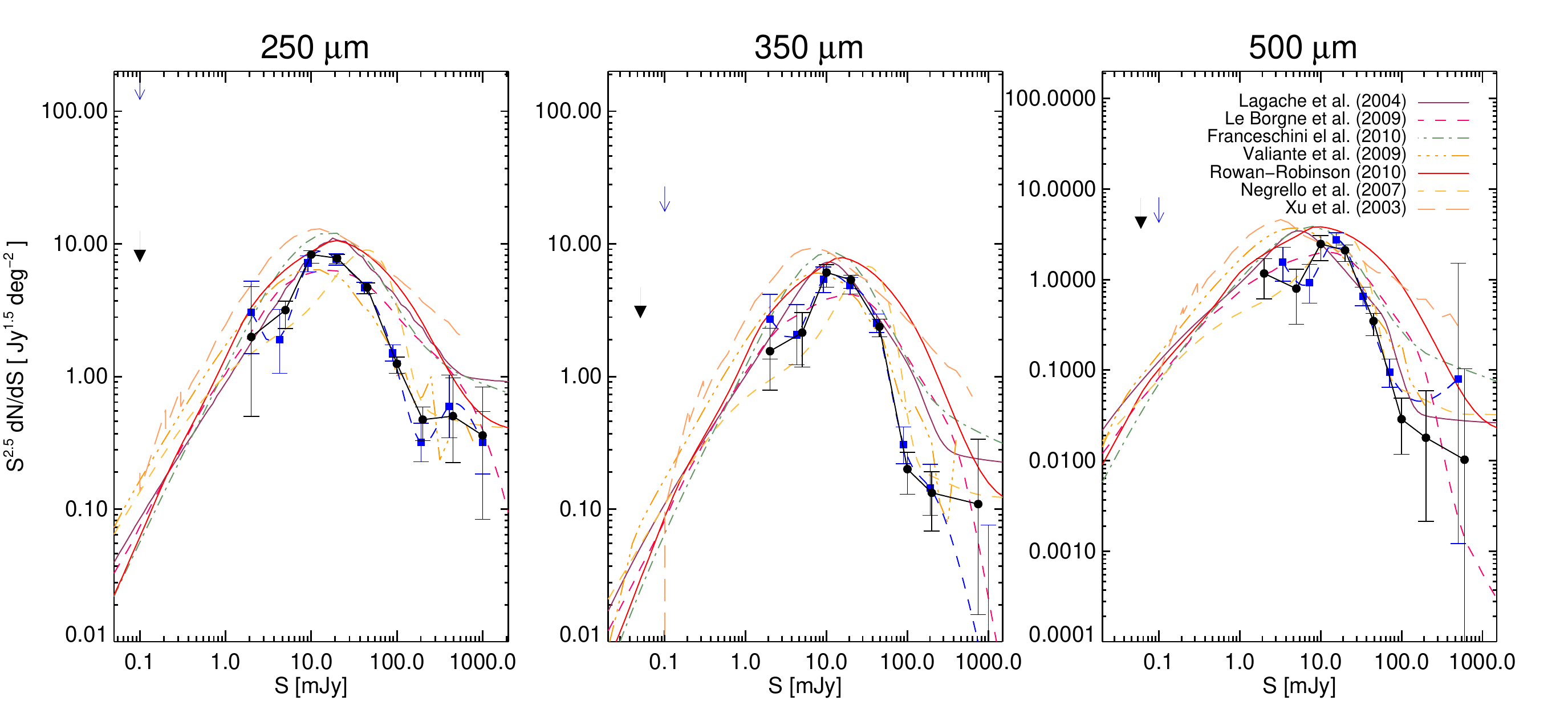}
\caption{Comparison of our Euclidean normalized differential number
  counts fits (as in figure~\ref{fig:brokpowoliver}, and not using the
  FIRAS prior) to a selection of models from the literature.  The
  error bars are the combined statistical and systematic errors.
\label{fig:modelcomp} }
\end{minipage}
\end{figure*}

Our fits are compared with a selection of literature models in
figure~\ref{fig:modelcomp}.  No currently available model entirely
fits our counts, especially when all three bands are considered.
There is a general discrepancy in the galaxy number count models in
that the theoretical models generically overpredict the number of
bright galaxies (in the several $\times$ 10 to several $\times$ 100
mJy range, limited at the upper end by uncertainties in the
\pd\ number counts) compared to the number counts from the
\pd\ analysis.  The best match overall across all three SPIRE bands is
given by the model of Valiante et al. (2009).

\begin{figure}
\centering
\includegraphics[width=9cm]{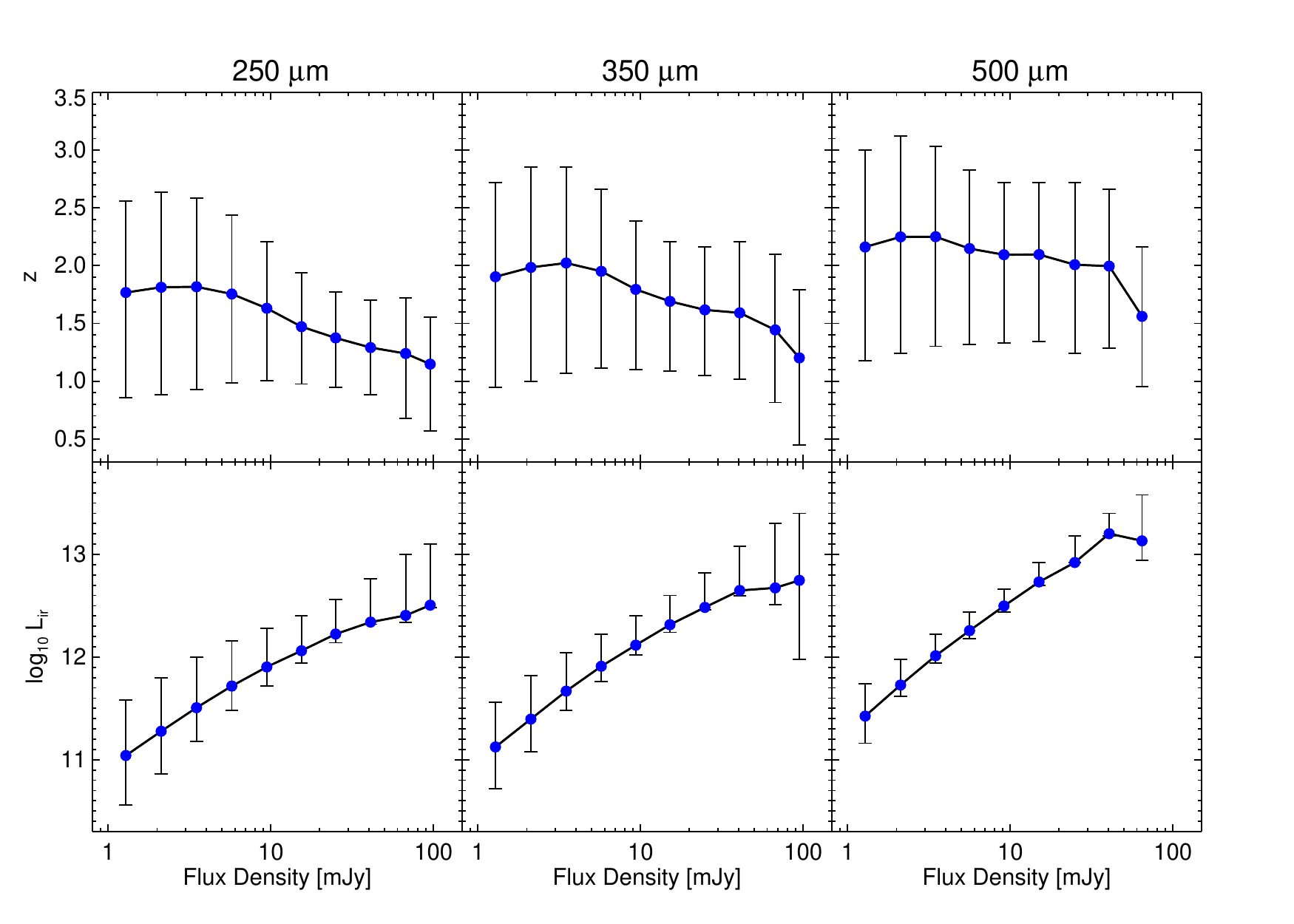}
\caption{The intrinsic redshifts (upper panels) and infrared
  luminosities (lower panels) of galaxies as a function of the
  observed flux densities in the SPIRE 250, 350, and 500~\mum\ bands
  (left, middle, and right panels, respectively) for the Lagache et
  al. models (Lagache et al. (2003); Fernandez-Conde et al. (2008)).
  The error bars give the interquartile range for each
  bin. \label{fig:galaxylum} }
\end{figure}

We interpret the discrepancy in the context of the theoretical models
of \cite{lag03} and \cite{fer08} (figure~\ref{fig:modelcomp}).  In
Figure \ref{fig:galaxylum} the redshifts and FIR luminosities of
galaxies are plotted versus their observed flux densities for the
\citet{fer08} simulations. The transitions from luminous to
ultra-luminous infra-red galaxies (LIRGs to ULIRGs, at $10^{12}
L_{\odot}$) with increasing observed flux densities occurs at
approximately 12, 6, and 3 mJy, in the 250, 350, and 500~\mum\ SPIRE
bands, respectively.  Thus, the discrepancy at the bright end likely
results from the presence of too many ULIRGs in the theoretical
models.  It should be noted that the intrinsic luminosities of the
underlying galaxy population that contribute to any given bin in
observed flux density depend on the redshift distributions and
spectral energy distributions; however, the very brightest galaxies
are likely either intrisically extremely luminous (ULIRGs or
brighter), low redshift, or strongly lensed.  There is a large dispersion in
redshifts represented by galaxies in each observed flux density bin,
with means in the range $z = 1 - 2$, with the average flux densities
only mildly inversely correlated with redshifts (due to the negative
$K$-correction).
 
In all three bands at and below 2 mJy, the \pd-derived number counts
are consistent with the theoretical galaxy number count models.  This
is not surprising because: (i) the theoretical galaxy number count
models are constrained not to overpredict the CFIRB, which arises, in
large part, from numerous faint galaxies; and (ii) the upper limits of
the lowest flux density knot in each band lies well above the
theoretical number counts models.

Another more subtle feature also seems apparent in the measured
counts.  The results from both fitting methods -- lending some
confidence to their credence -- have depressions at the third-lowest
flux density knots with respect to the theoretical number count models
at low-to-moderate significance (depending on the theoretical model),
which are exclusively concave down in this range (a few mJy to a few
times 10 mJy).  The stacking analysis of \citet{Bethermin:2010a} is
not deep enough at 250 or 350~\mum\ to verify this feature, although
the turnover in $S^{2.5} dN/dS$ from the peak at approximately 10 mJy
downward is clear at 250 \mum .  At 500~\mum, the stacking analysis
does not display the depression.  This flux density range is
approximately at the confusion limit (where the flux density is equal
to the confusion noise, $\sigma_{\mathrm{conf}} = 6$ mJy) and multiple
galaxies contribute to the flux density in each beam.  Thus, referring
to the \citet{fer08} models (figure~\ref{fig:galaxylum}), this flux
density range corresponds to the transition from ULIRGS to LIRGs,
suggesting that LIRG number counts may also be overrepresented by the
theoretical galaxy number counts models.

\section{Conclusions}
\label{sec:conclusions}
\nobreak 
We have measured the differential galaxy number counts from {\it
  Herschel}-SPIRE Science Demonstration Phase HerMES observations at
250, 350, and 500 \mum\ using \pd\ techniques and two simple
parametric models.  The number counts were measured down to 2 mJy,
approximately a factor of 3 below the 1 $\sigma$ confusion noise.  We
find that $64 \pm 14$ per cent of the measured CFIRB is accounted for
by point sources at 250 \mum\ , falling to $43 \pm 12$ per cent at 500
\mum .  The errors on the fraction of the CFIRB accounted for by these
sources are now dominated by those of the FIRAS measurement.  However,
because of the remaining fraction not accounted for by our fits, this
is still not a competitive method for measuring the total CFIRB.  We
find clear evidence of breaks in the slope of the differential number
counts at approximately 10--20 mJy in all bands, which have been
hinted at by previous analyses.

Where they overlap, our fits agree well with other {\it Herschel}
results.  Comparing with a selection of literature models, however, we
find that no model entirely reproduces our observed number counts.  As
found by \citet{Oliver:2010b} and \citet{Clements:2010}, most
published models significantly over-predict the number of bright
sources at these wavelengths and have shallower slopes.  We find
somewhat better agreement at fainter fluxes, at or below the break,
but the agreement is still not perfect.

Our main systematic uncertanties arise from our understanding of the
SPIRE beams.  We find that a high-pass filter is effective in removing
the signature of clustering from our counts, but in the future it may
be preferable to attempt to directly marginalize over clustering using
simple models.  

These observations represent only $\sim 60$ hours of the 900 hours of
observations that HerMES will ultimately obtain (although not all of
these are with SPIRE).  The final dataset will cover a wide range of
depths and areas.  This will significantly increase our ability to constrain
$dN/dS$.  Having a number of well-seperated deep
fields will also allow a direct measurement of sample variance.

{\it Acknowledgments.}  The authors would like to thank Guillaume
Patanchon and Phil Maloney for many useful discussions.  J.~Glenn and
A.~Conley acknowledge support from NASA Herschel GTO grant 1394366,
sponsored by the Jet Propulsion Laboratory.  SPIRE has been developed
by a consortium of institutes led by Cardiff Univ.\ (UK) and including
Univ.\ Lethbridge (Canada); NAOC (China); CEA, LAM (France); IFSI,
Univ.\ Padua (italy); IAC (Spain); Stockholm Observatory (Sweden);
Imperial College London, RAL, UCL-MSSL, UKATC, Univ.\ Sussex (UK);
Caltech, JPL, NHSC, Univ.\ Colorado (USA).  This developement has been
supported by national funding agencies: CSA (Canada); NAOC (China);
CEA, CNES, CNRS (France); ASI (Italy); MCINN (Spain); SNSB (Sweden);
STFC (UK); and NASA (USA).  The data presented in this paper will be
released through the {\it Herschel} Database in Marseille, HeDaM
(http://hedam.oamp.fr/HerMES).


\begin{thebibliography}{}
\bibitem[\protect\citeauthoryear{Barcons}{1992}]{barc92} Barcons, X.\
 1992, ApJ, 396, 460
\bibitem[\protect\citeauthoryear{Barcons}{1994}]{Barcons:1994} Barcons, X.\
 et al.\ 1994, MNRAS, 268, 833
\bibitem[\protect\citeauthoryear{Berta}{2010}]{Berta:2010} Berta, S.\
 et al.\ 2010, A\&A, 518, 30
\bibitem[\protect\citeauthoryear{B{\'e}thermin et al.}{2010a}]{Bethermin:2010a}
 B{\'e}thermin, M.\ et al.\ 2010a, A\&A, 512, 78
\bibitem[\protect\citeauthoryear{B{\'e}thermin et al.}{2010b}]{Bethermin:2010b}
 B{\'e}thermin, M.\ et al.\ 2010b, A\&A, 516, 43
\bibitem[\protect\citeauthoryear{Blain et al.}{2004}]{Blain:2004}
 Blain, A.\ et al.\ 2004, ApJ 611, 725
\bibitem[\protect\citeauthoryear{Chapin et al.}{2010}]{Chapin:2010}
 Chapin, E.~L.\ et al.\ 2010, ArXiv e-print 1003.2647
\bibitem[\protect\citeauthoryear{Clements et al.}{2010}]{Clements:2010}
 Clements, D.~L.\ et al.\ 2010, A\&A, 518, 8
\bibitem[\protect\citeauthoryear{Cooray et al.}{2010}]{Cooray:2010}
 Cooray, A.\ et al.\ 2010, A\&A, 518, 22
\bibitem[\protect\citeauthoryear{Condon}{1974}]{con74} Condon, J.~J. 1974, ApJ,
  188, 279
\bibitem[\protect\citeauthoryear{Devlin et al.}{2009}]{Devlin:2009}
 Devlin, M.~J.\ et al.\ 2009, Nature, 458, 737
\bibitem[\protect\citeauthoryear{Dole et al.}{2006}]{dol06} 
 Dole, H., et al.\ 2006, A\&A, 451, 417
\bibitem[\protect\citeauthoryear{Dowell et al.}{2010}]{Dowell:2010}
  Dowell, C.~D.\ et al.\ 2010, Space Telescopes and Instrumentation
  2010: Optical, Infrared, and Millimeter Wave, Proc. SPIE 7731, in
  press
\bibitem[\protect\citeauthoryear{Dunkley et al.}{2005}]{Dunkley:05}
 Dunkley,~J.\ et al.\ 2005, MNRAS, 356, 925
\bibitem[\protect\citeauthoryear{Fernandez-Conde et al.}{2008}]{fer08}
  Fernandez-Conde, N. et al.\ 2008, A\&A, 481, 885
\bibitem[\protect\citeauthoryear{Fixsen et al.}{1998}]{Fixsen:1998} Fixsen,
  D.J., Dwek, E., Mather, J.C., Bennett, C.L., \& Shafer, R.A. 1998,
  ApJ, 508, 123
\bibitem[\protect\citeauthoryear{Franceschini et al.}{2010}]{Franceschini:2010}
 Franceschini, A.\ et al.\ 2010, A\&A in press
\bibitem[\protect\citeauthoryear{Gelman \&\ Rubin}{1992}]{Gelman:1992}
 Gelman, A.\ and Rubin, D.~B.\ 1992, Statistical Science, 7, 457
\bibitem[\protect\citeauthoryear{Griffin et al.}{2010}]{Griffin:2010}
  Griffin, M.\ et al.\ 2010, A\&A, 518, 3
\bibitem[\protect\citeauthoryear{Guiderdoni et al.}{1997}]{guid97}
  Guiderdoni, B., Bouchet, F.R., Puget, J.-L., Lagache, G., \& Hivon,
  E 1997, Nature, 390, 257
\bibitem[\protect\citeauthoryear{Hughes et al.}{1998}]{Hughes:1998}
 Hughes, D.~H.\ et al.\ 1998, Nature, 394, 241
\bibitem[\protect\citeauthoryear{Lagache et al.}{2003}]{lag03}
  Lagache, G., Dole, H, \& Puget, J.-L.\ 2003, MNRAS, 338, 555
\bibitem[\protect\citeauthoryear{Lagache et al.}{2007}]{Lagache:2007}
  Lagache, G.\ et al.\ 2007, ApJ, 665, 89
\bibitem[\protect\citeauthoryear{Laurent et al.}{2005}]{laur05}
  Laurent, G.T.\ et al.\ 2005, ApJ, 623, 742
\bibitem[\protect\citeauthoryear{Le~Borgne et al.}{2009}]{LeBorgne:2009}
 Le~Borgne, D., et al.\ 2009, A\&A, 504, 727
\bibitem[\protect\citeauthoryear{Lewis \& Bridle}{2002}]{Lewis:02}
 Lewis,~A. and Bridle,~S.\ 2002, PhRVD, 66, 103511
\bibitem[\protect\citeauthoryear{MacKay}{2002}]{MacKay:02} MacKay,
  D.~J.~C., 2002, Information Theory, Inference, and Learning
  Algorithms, Cambridge University Press
\bibitem[\protect\citeauthoryear{Maloney et al.}{2005}]{mal05}
  Maloney, P.R., et al.\ 2005, ApJ, 635, 1044
\bibitem[\protect\citeauthoryear{Marsden et al.}{2009}]{mar09} Marsden,
  G., et al.\ 2009, ApJ, 707, 1729
\bibitem[\protect\citeauthoryear{Negrello et al.}{2007}]{Negrello:2007}
 Negrello, M.\ et al.\ 2007, MNRAS, 377, 1557
\bibitem[\protect\citeauthoryear{Nguyen et al.}{2010}]{Nguyen:2010}
  Nguyen, H.~T.\ et al.\ 2010, A\&A, 518, 5
\bibitem[\protect\citeauthoryear{Oliver et al.}{1997}]{Oliver:1997}
  Oliver, S.~J.\ et al.\ 1997, MNRAS, 289, 471
\bibitem[\protect\citeauthoryear{Oliver et al.}{2010}]{Oliver:2010b}
  Oliver, S.~J.\ et al.\ 2010, A\&A, 518, 21
\bibitem[\protect\citeauthoryear{Patanchon et al.}{2009}]{pat09}
  Patanchon, G.\ et al.\ 2009, ApJ, 707, 1750
\bibitem[\protect\citeauthoryear{Pilbratt et al.}{2010}]{Pilbratt:2010}
  Pilbratt, G.\ et al.\ 2010, A\&A, 518, 1
\bibitem[\protect\citeauthoryear{Puget et al.}{1996}]{pug96} Puget, J.-L.\
  et al.\ 1996, A\&A, 308, L5
\bibitem[\protect\citeauthoryear{Rowan-Robinson}{2001}]{RR01}Rowan-Robinson,
  M. 2001, ApJ, 549, 745
\bibitem[\protect\citeauthoryear{Rowan-Robinson}{2009}]{RR09}Rowan-Robinson,
  M. 2009, MNRAS, 394, 117
\bibitem[\protect\citeauthoryear{Saunders}{1990}]{Saunders:90} Saunders, W.\
  1990, MNRAS, 242, 318
\bibitem[\protect\citeauthoryear{Scheuer}{1957}]{scheuer57} Scheuer,
  P.~A.~G. 1957, Proc.~Camb.~Philos.~Soc., 53, 764
\bibitem[\protect\citeauthoryear{Scheuer}{1974}]{scheuer74} Scheuer,
  P.~A.~G. 1974, MNRAS, 166, 329
\bibitem[\protect\citeauthoryear{Scott et al.}{2002}]{scott02} Scott,
  S.~E., et al.\ 2002, MNRAS, 331, 817
\bibitem[\protect\citeauthoryear{Scott et al.}{2010}]{Scott:2010} Scott,
  K.~S., et al.\  2010, MNRAS, 405, 2260
\bibitem[\protect\citeauthoryear{Smail et al.}{2002}]{smail} Smail,
  I., Ivison, R.J., Blain, A.W., \& Kneib, J.-P. 2002, MNRAS, 331,
  495
\bibitem[\protect\citeauthoryear{Somerville et al.}{2004}]{Somerville:2004}
 Somerville, R.~S.\ et al.\ 2004, ApJ, 600, 171
\bibitem[\protect\citeauthoryear{Swinyard et al.}{2010}]{Swinyard:2010}
  Swinyard, B. et al.\ 2010, A\&A, 518, 4
\bibitem[\protect\citeauthoryear{Takeuchi \& Ishii}{2004}]{ti04}
  Takeuchi, T.~T., \& Ishii, T.T. 2004, ApJ, 604, 40
\bibitem[\protect\citeauthoryear{Valiante et al.}{2009}]{Valiante:2009}
 Valiante, E.\ et al.\ 2010, ApJ, 701, 1814
\bibitem[\protect\citeauthoryear{Viero et al.}{2009}]{Viero:2009}
  Viero, M.~P.\ et al.\ 2009, ApJ, 707, 1766
\bibitem[\protect\citeauthoryear{Wei{\ss} et al.}{2009}]{Weiss:2009}
 {Wei{\ss}}, A.\ et al.\ 2009, ApJ, 707, 1201
\bibitem[\protect\citeauthoryear{Xu et al.}{2003}]{Xu:2003}
 Xu, C.~K.\ et al.\ 2003, ApJ, 587, 90
\bibitem[\protect\citeauthoryear{Zemcov et al.}{2010}]{Zemcov:2010}
 Zemcov, M.\ et al.\ 2010, submitted to ApJ
\end{thebibliography}
\end{document}